\documentstyle[draft]{mn}

\newcommand{\etal}{et~al.\ }

\newcommand{\mum}{$\,\mu$m}
\newcommand{\pho}{\phantom{1}}
\newcommand{\phm}{\phantom{-}}
\newcommand{\phg}{\phantom{{<}\,}}

\input epsf
\input psfig.sty
\input epsfig.sty

\begin{document}

\def\gs{\mathrel{\raise0.35ex\hbox{$\scriptstyle >$}\kern-0.6em
\lower0.40ex\hbox{{$\scriptstyle \sim$}}}}
\def\ls{\mathrel{\raise0.35ex\hbox{$\scriptstyle <$}\kern-0.6em
\lower0.40ex\hbox{{$\scriptstyle \sim$}}}}

\journal{Preprint UBC-COS-00-04, astro-ph/0009067}
\title[Submillimetre sources in rich cluster fields]
{Submillimetre sources in rich cluster fields -- source counts,
redshift estimates, and cooling flow limits}

\author[S.\,C.\ Chapman et al.]
       {Scott\,C.\ Chapman,$^{1}$ 
	Douglas\ Scott,$^{2}$ Colin\ Borys,$^{2}$ and 
	Gregory\,G.\ Fahlman$^{2,3}$
	\vspace*{1mm}\\
        $^1$Observatories of the Carnegie Institution of Washington,
        Pasadena, CA 91101,~~U.S.A.\\
        $^2$University of British Columbia, Department of Physics \& Astronomy, 
         Vancouver BC V6T 1Z4, Canada\\
	$^3$Canada-France-Hawaii Telescope, 
                  Kamuela, Hawaii 96743,~~~USA
        }

\date{Accepted ... ;
      Received ... ;
      in original form ...}

\pagerange{000--000}

\maketitle

\begin{abstract}
Recent submillimetre surveys have revealed a population of dusty, high redshift
sources of great cosmological significance for understanding dust-enshrouded
star formation in distant galaxies, and for determining the origin of the
far-IR background. 
In this paper, we analyze nine rich cluster fields mapped at 850\mum\ and
450\mum\ with the SCUBA array on the James Clerk Maxwell telescope.
Lensing models of the clusters are developed in order to derive 
accurate source counts for our sample.  
VLA maps of the same clusters are used to help constrain the redshift 
distribution of our SCUBA detections.
Implications for high redshift
galaxies and for the far-IR background are discussed. 
We also provide limits on distributed dust produced by cooling flows
in these clusters.
\end{abstract}


\section{Introduction}
The submillimetre waveband has recently  become an invaluable tool for
investigating  the properties of  high redshift galaxies. An early result
using the Sub-millimetre Common User Bolometer Array 
(SCUBA -- Holland et al. 1999) by
Smail, Ivison and Blain (1997) showed  that a  much larger
population  of dusty, high redshift  galaxies  existed than previously thought.  
The existence of these dusty, sub-mm galaxies at relatively high
redshifts have now been confirmed in several surveys (Hughes et al.~1998;
Barger et al.~1998, 1999a; Eales et al.~1999; Lilly et al.~1999a;
Blain et al.~1999). Such sub-mm sources 
could be of great cosmological significance -- they may comprise a substantial
fraction of the Far-Infrared Background 
(FIRB -- Puget et al.~1996, Hauser et al.~1998, Lagache et al.~1999).
Additionally, they could be an important source of CMB anisotropy at
arcsecond scales (Scott \& White 1999), and certainly the dominant source
of sky fluctuations at
${\sim}\,1$mm, at least out of the galactic plane (Hughes et al.~1998, 
Borys, Chapman \& Scott 1999).
Our understanding of the star-formation history of galaxies
depends critically on the properties of these sub-mm sources:
this population is responsible for a substantial fraction of the total
star formation output in the early Universe and holds important information 
about how galaxies formed and evolved
(e.g.~Blain \& Longair 1993, Guiderdoni et al.~1997, Blain et al.~1998).
They appear to be the
high redshift counterparts of well-studied nearby ultra-luminous IR galaxies
(Sanders \& Mirabel~1996),
and can be plausibly intrepretted as the formation sites
of galactic spheroids through mergers of gaseous disk systems (e.g.~Sanders
1999, Lilly et al.~1999a).

Much  work  has gone  into identifying just what
types of galaxies  make  up this  sub-mm  bright population, and  what
mechanisms may be responsible for their rest-frame far-IR emission
\cite{Smaetal,Hugetal,Ivietal,Liletal,Frayer98,Frayer99,Frayer00,Chapman99,Iv2000,ChapLBG,MAMBO,Gear00}.
However, the numbers of sub-mm source detections 
are still relatively small.  In an effort to increase the statistics and
broaden the source count baseline, we have undertaken a sub-mm survey
including both blank sky (Borys et al.~in preparation)
and cluster fields.  Our cluster sample is
complementary to that of Smail, Ivison, and Blain (1997, 1998)
who pioneered this approach using SCUBA on the JCMT.
These cluster fields benefit from the lensing by the foreground cluster
mass at $z\,{<}\,1$, enhancing sensitivity to the dusty galaxies at higher
redshifts.                                                                
 
Our cluster sample consists of nine clusters at redshifts 0.2--0.8,
originally selected to search for both distributed cool dust
(cooling flows -- see Edge et al.~1999) and the Sunyaev-Zel'dovich (SZ)
effect (Birkinshaw 1999; Andreani et al.~1999).
Although we found no convincing evidence for distributed emission, 
our SCUBA maps have revealed significant source detections,
as expected from the counts from other surveys (e.g. Blain \etal 1999,
Barger \etal 1999b). This further demonstrates that
sub-mm surveys provide an efficient means of identifying a
population of dusty AGN and/or star-forming galaxies, possibly at very 
high redshifts.
In addition,  a few weak sources appear coincident with radio-bright
central galaxies in the clusters, possibly associated with the
strong cooling flows suggested by the x-ray profiles. 
In this paper, we describe possible sources found in the cluster fields, 
and present the resulting 
source counts. Existing VLA maps are used to place limits on the redshift
distribution of the sample. We conclude by discussing the implications for
high redshift galaxies and the FIRB.

\section{Observations}

\subsection{JCMT data}
Nine clusters were observed with the  SCUBA instrument
on the James Clerk Maxwell Telescope, resulting
in 17 significant source detections greater than 3$\sigma$.
During 4 observing runs throughout 1998 and 1999, we
operated the  91 element Short-wave
array at 450\mum, and the 37  element Long-wave array at 850\mum\
simultaneously, in JIGGLE mode, and additionally the single photometry 
pixel at 1350\mum\ for some sources, 
giving half-power beam  widths of 7.5, 14.7, and 21.0 arcsec respectively.    
Additionally, 850/450\mum\ PHOTOMETRY mode followup on 3 of the sources was 
performed, confirming their existence and flux densities.
The central pixel  of SCUBA   was fixed on  the source, as defined by
the centroid of the {\sl ROSAT} imaging (Crawford et al.~1999).
For mapping, the standard 64-point jiggle pattern
was employed  to fully sample the 450 and 850\mum\ arrays. The effective
integration times (time spent on and off source derived from the number
of 64 point integrations obtained, which excludes overheads) 
are presented in Table~1 for each cluster.
For 1350\mum\ photometry we used the 9-point jiggle pattern
to reduce the impact of  pointing errors by averaging the
source signal over a slightly larger area than the beam, resulting in
greater photometric  accuracy.    
Whilst jiggling, the  secondary  was
chopped at 7.8125\,Hz with chop throws between 40 and 120 arcsec in azimuth, 
with smaller throws used under worse  
atmospheric conditions or when the projected core radius for a 
cluster was smaller.

Pointing was checked hourly on blazars and a sky-dip was
performed to measure the atmospheric
opacity directly. The RMS pointing errors were below 2 arcsec, while  the
average atmospheric zenith  opacities at 450\mum\,, 850\mum\  and
1350\mum\ were fairly stable and generally quite good  with $\tau_{850}$ 
ranging from 0.15 to 0.36.  Some short time-scale  variations,
presumably due to the atmosphere, caused some parts of the data-set to be
noisier, and these noisiest sections were excluded. 

The data were reduced using the  Starlink package SURF (Scuba User
Reduction Facility, Jenness et al.~1998).
Spikes were first carefully rejected from the 
data, followed by correction for atmospheric opacity
and sky subtraction using the median of all the array pixels, except for
obviously bad pixels (the 1350\mum\ pixel currently
has no provision  for subtracting sky  variations using the other
wavelength pixels). 
We then weighted each pixel by its timestream inverse variance,
relative to the central pixel. Multiple scans on the same cluster field,
and sky rotation, ensure that (in most cases) 
each point of the map is covered by many bolometers. 
Even with the inverse variance weighting, excessively bad pixels still 
appeared to degrade the map, and were therefore 
clipped from the data prior to rebinning into
the final maps.
The data were  then calibrated against standard planetary or 
compact \hbox{H\,{\sc ii}~} region sources, observed  during the
same  nights as the clusters. 
Uncertainties at 850\mum\  and 1350\mum\ are estimated to be around
10 per cent, while at 450\mum\ they could be as high as 25 per cent.
Note that calibration uncertainties do not affect the S/N level for our
detections, only the overall flux densities reported.

\subsection{Archive VLA data}

VLA maps at 4.9\,GHz in the C configuration were obtained for five clusters 
in our sample courtesy of A.~Edge. The effective resolution in the images
is 5 arcsec.  More details of the reduction and analysis of similar maps 
are given in Edge et al.~(1999a).
Typical 1$\sigma$ sensitivities lie around $60\,\mu$Jy.
In addition, an archive 1.4\,GHz map for Abell\,2219 was obtained courtesy of
F.~Owen, reaching a 1$\sigma$ limit of ${\sim}\,39\,\mu$Jy per 2 arcsec beam.

For the remaining clusters, VLA data presented in Stocke et al.~(1999)
at 1.4\,GHz in the B configuration, was used to set limits and identify 
counterpart positions.
Typical 1.4\,GHz sensitivities in these maps are 65\,$\mu$Jy RMS.

\setcounter{figure}{0}
\begin{figure*}
\begin{minipage}{170mm}
\begin{center}
\epsfig{file=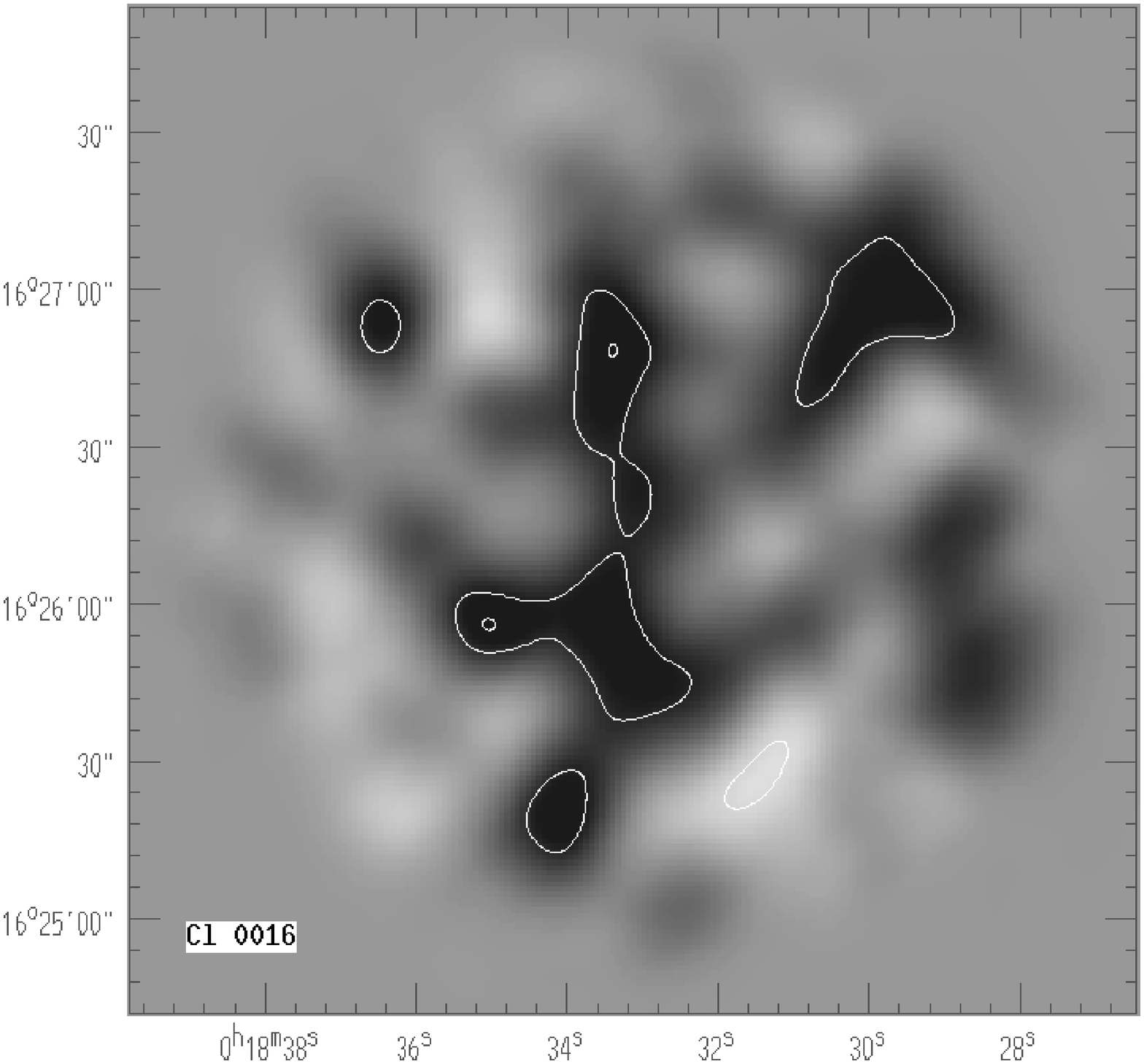, width=8.025cm, angle=0}
\epsfig{file=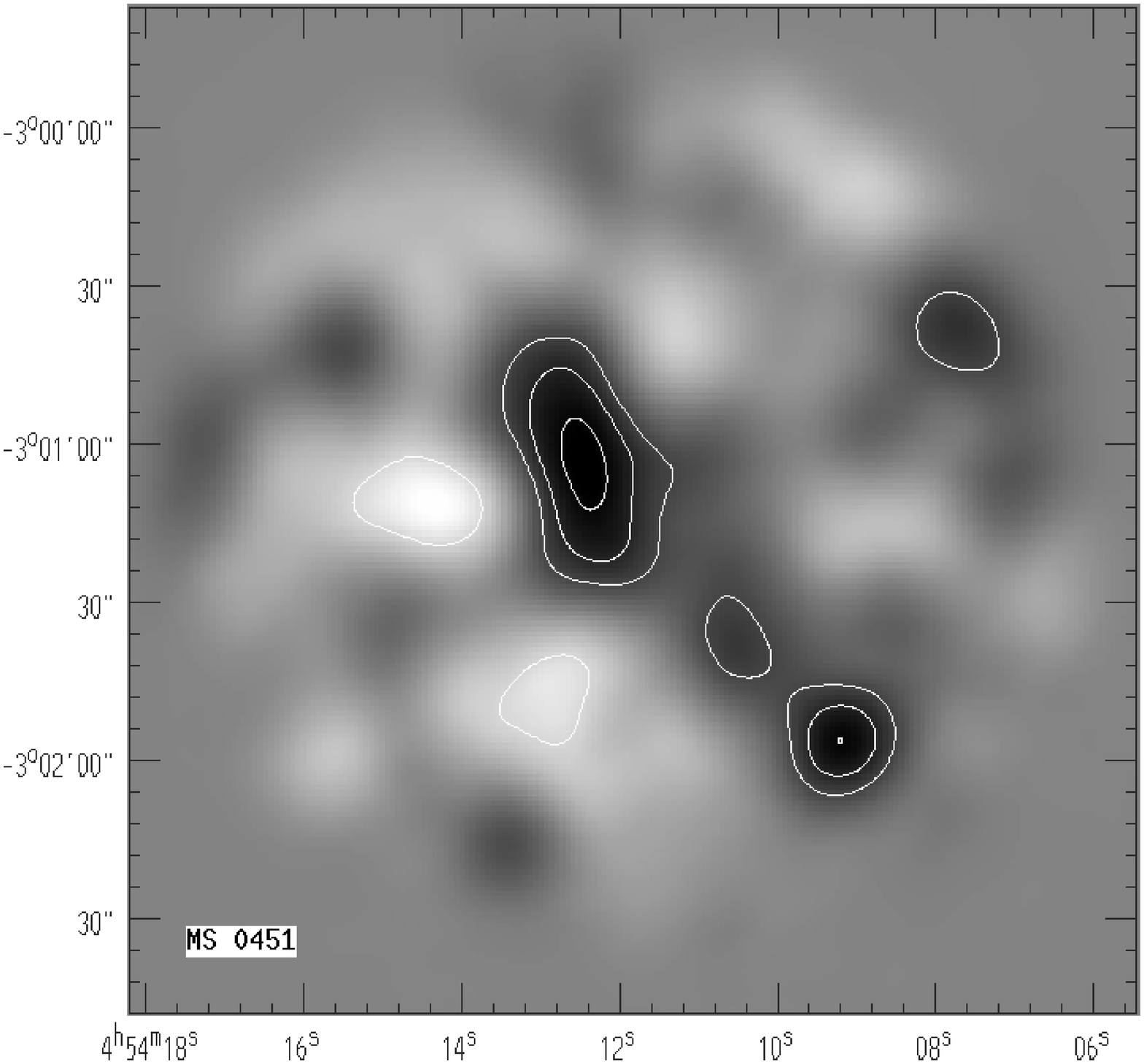, width=8.025cm, angle=0}
\epsfig{file=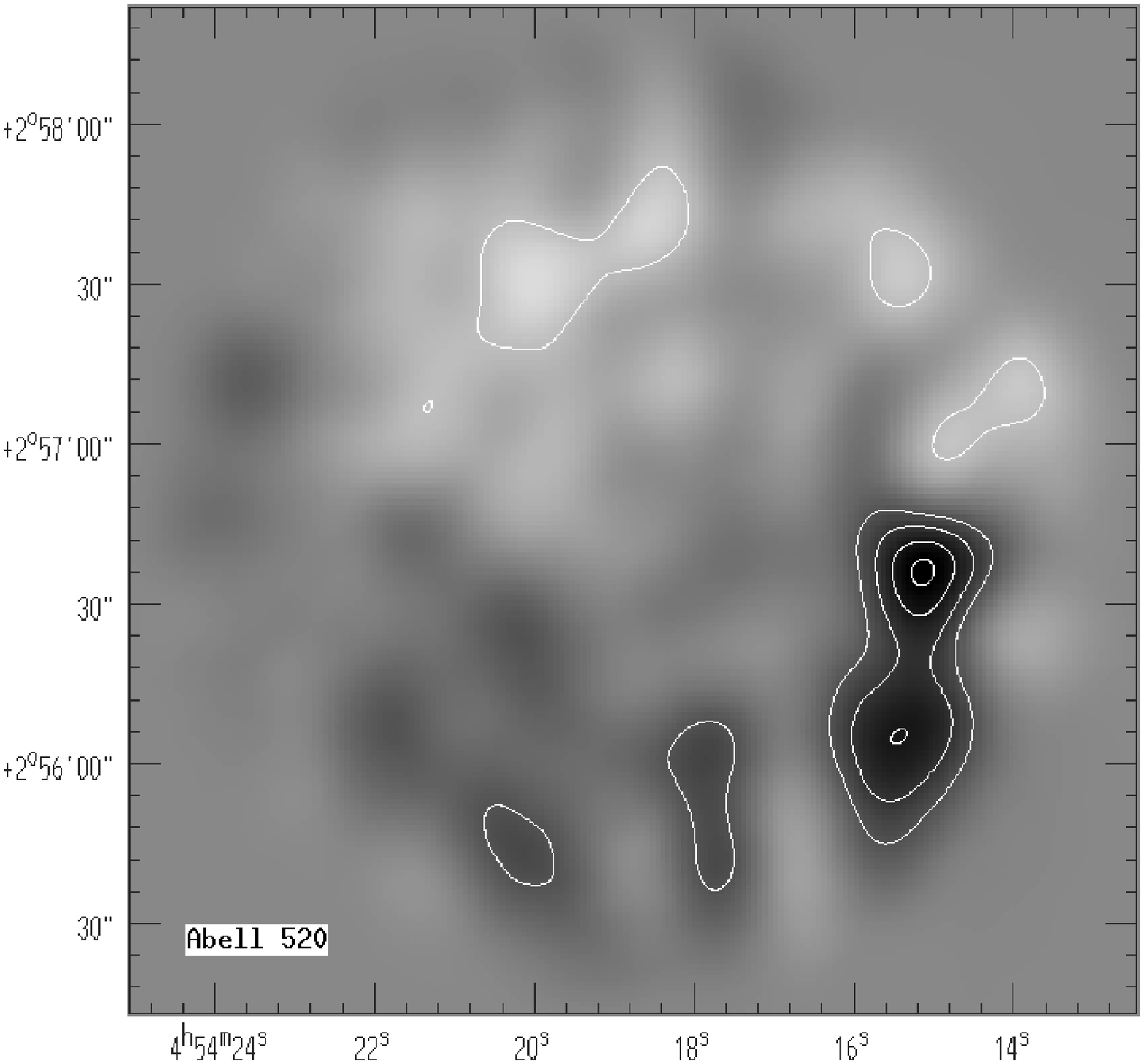,  width=8.025cm, angle=0}
\epsfig{file=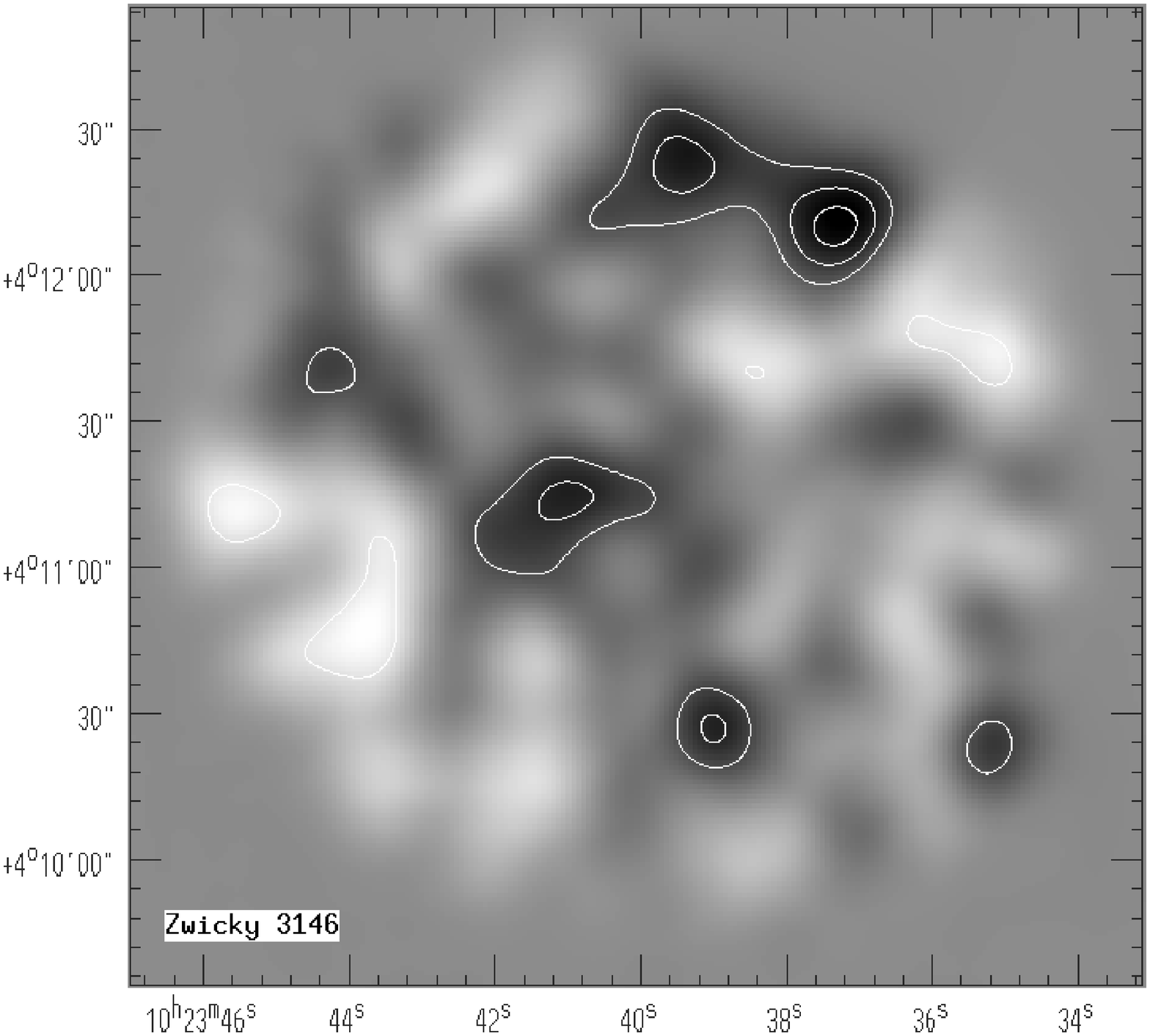, width=8.025cm, angle=0}
\end{center}
\caption{Cluster field images at 850\mum, convolved with the SCUBA 14.7 arcsec
beam. 
Dark shading is positive flux density, while lighter shading is negative.
All contour levels indicate the
number of sigma above the background noise estimated from the central
regions of the beam-convolved
images, starting at 2$\sigma$, with intervals
of 1$\sigma$. These noise levels correspond approximately to the tabulated
values (Tables 1\&2), which were derived from the raw data. 
Top row: Cl\,0016+16 (left); MS\,0451-03 (right). 
Bottom row: Abell\,520 (left);  Zwicky\,3146 (right).
}
\label{f-clusters}
\end{minipage}
\end{figure*}

\begin{figure*}
\begin{minipage}{170mm}
\begin{center}
\epsfig{file=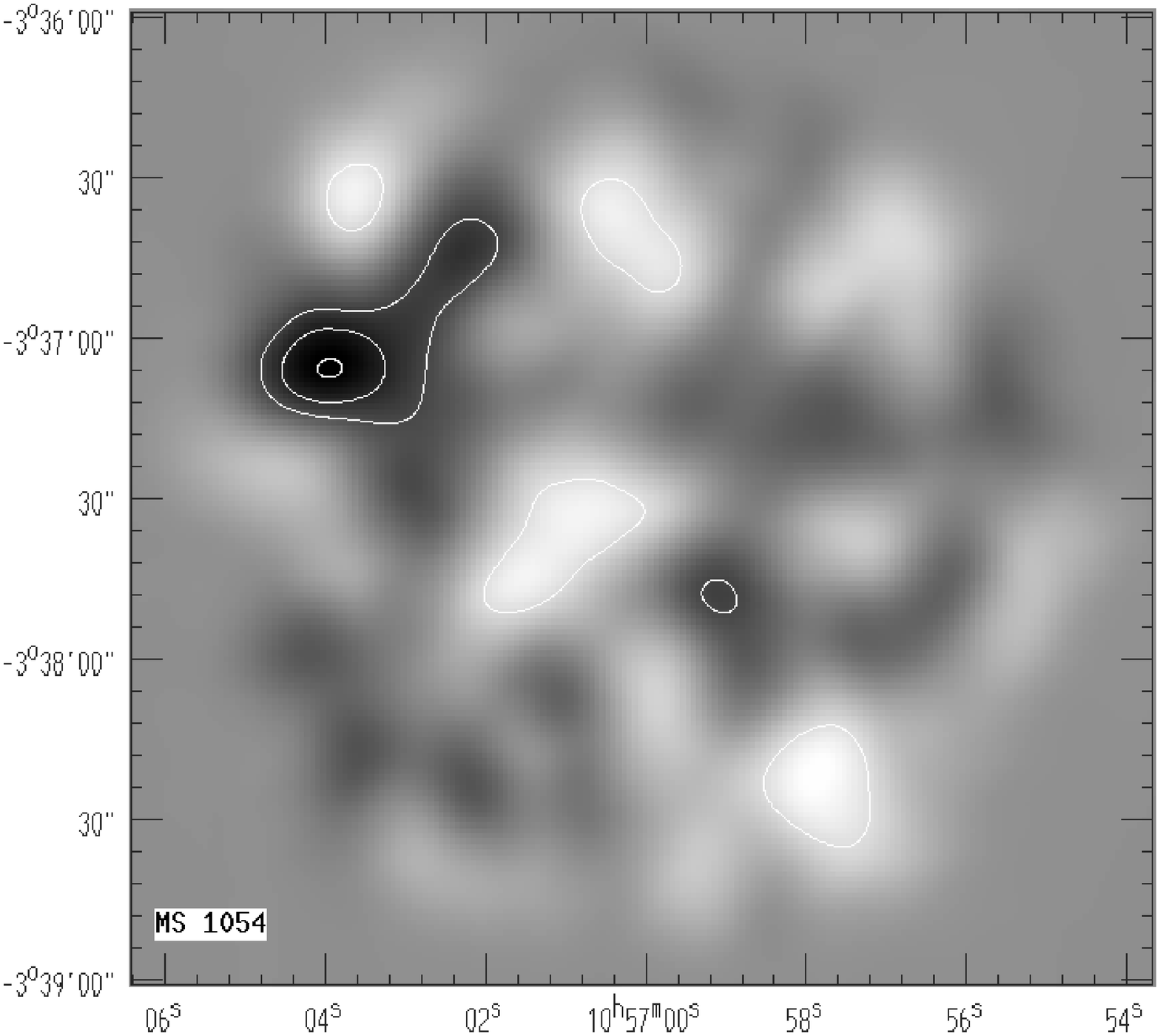, width=8.025cm, angle=0}
\epsfig{file=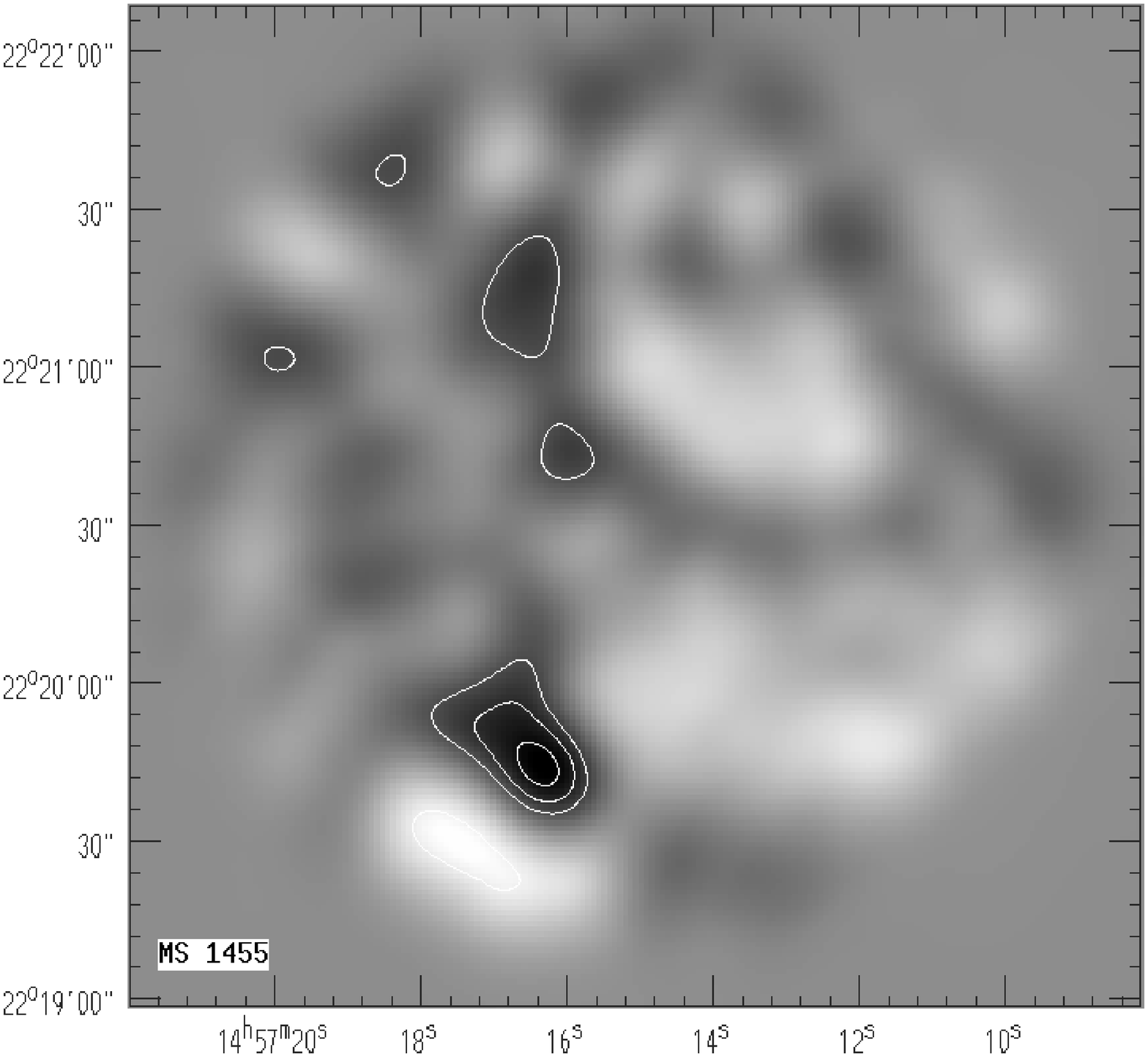, width=8.025cm, angle=0}
\epsfig{file=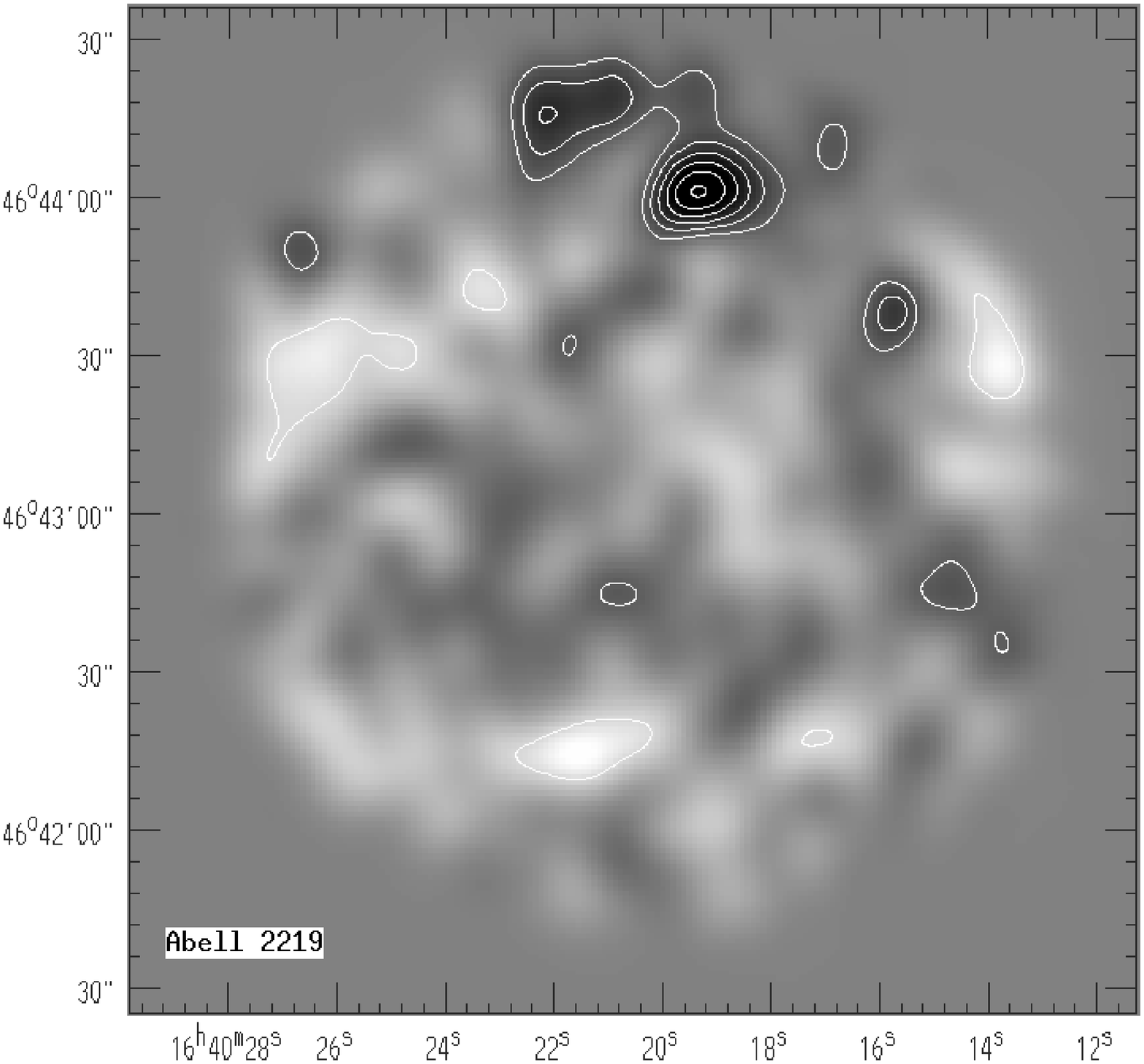, width=8.025cm, angle=0}
\epsfig{file=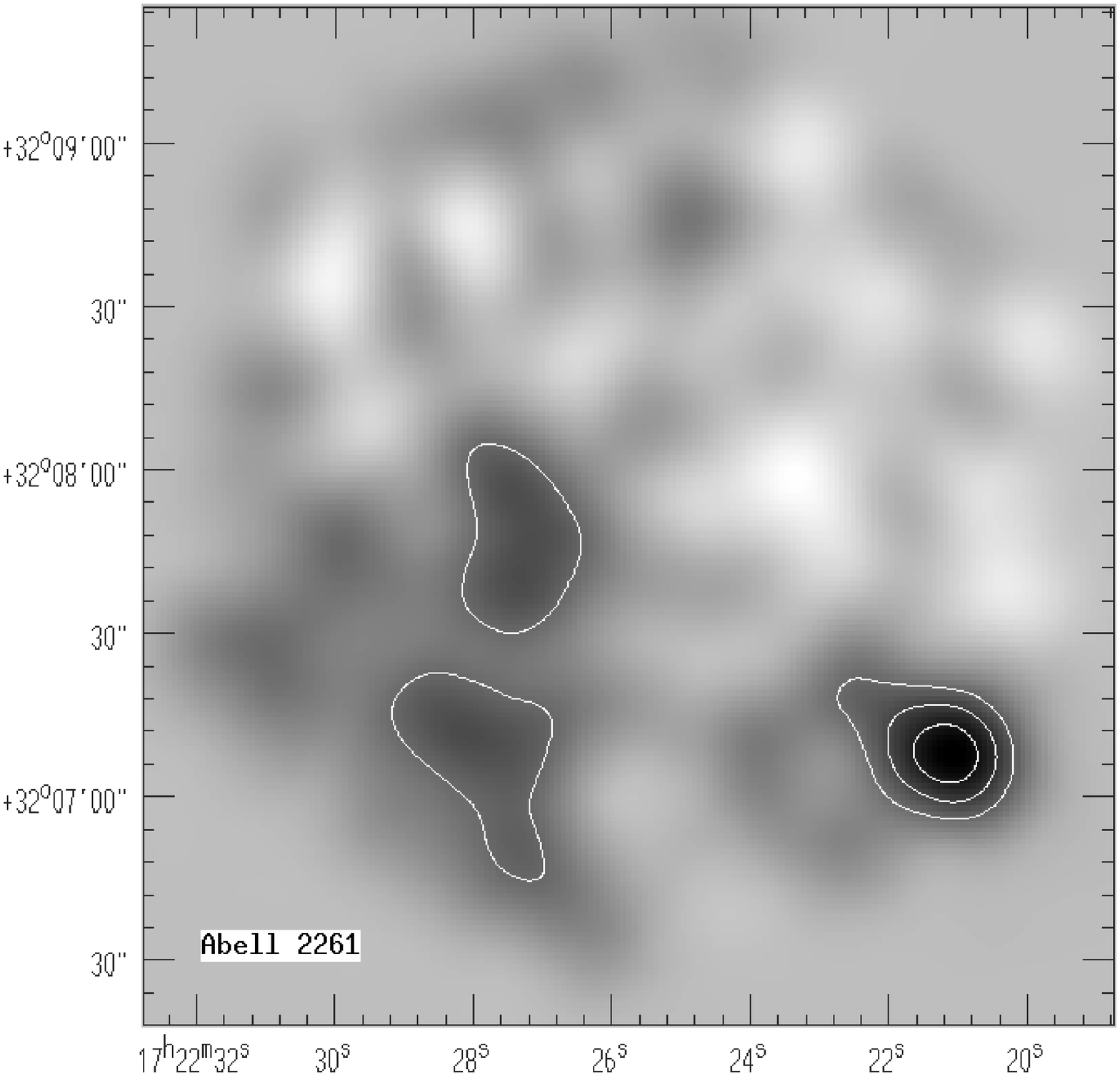, width=7.525cm, angle=0}
\end{center}
\caption{SCUBA maps at 850\mum\ continued. Contours are the same as Fig.~1.
Top row: MS\,1054-03 (left); MS\,1455+22 (right).
Bottom row: Abell\,2219 (left); Abell\,2261 (right).
}
\label{f-clusters2}
\end{minipage}
\end{figure*}                 

\section{Source Catalogue}

The 850\mum\ SCUBA maps for the clusters showing 
significant source detections are presented in Figs.~\ref{f-clusters} 
and \ref{f-clusters2}.
Any real objects in our maps are expected to be unresolved and will 
typically appear as a positive point spread function 
with FHWM ${\sim}\,15$~arcsecs at 850\mum\,.
For most of our maps, we do not expect to see negative signatures for our 
sources, as the individual scans coadded in the map have 
different chop throws and directions, diluting the effect of off-beams.
We also chopped out of the field in most cases.
The edges of the maps incur a  rapid
rise in noise level due to incomplete 
sampling of the sky for the outer bolometers of the SCUBA array.
Eales et al. (1999) simulated the incomplete sampling at the field
edge and found that any spurious sources in Monte Carlo simulations tended
to appear in this region.
We have therefore clipped the 14 arcsec edge regions in our maps
where the noise level increases by a factor $\ga$\,2 times that at the field
center.

As the signal-to-noise ratio (S/N) is low for individual pixels, we use a 
template fitting method for source detection to achieve a maximum 
gain in S/N, similar to that used in Eales et al.~(1999). The cluster maps are 
convolved with the beam map obtained by imaging Mars, and all peaks greater
than $3\sigma$ are extracted as potential sources.
We checked the convolved maps carefully against the raw data to ensure that no
residual bad pixels were identified as sources.  We also note that there are no
strong negative sources in our maps.

Sensitivity limits for deep SCUBA integrations have now been fairly well
characterized (Smail \etal 1997, Hughes \etal 1998, Eales \etal 1998, Barger
\etal 1999b).
But at these low S/N levels, we must be careful to consider possible
systematic noise effects.  There are currently five
identified components which contribute to the overall SCUBA noise level:
(i) photon noise from the sky;
(ii) telescope and instrumental noise;
(iii) blanked-off detector Johnson, phonon and amplifier noise;
(iv) excess photon noise from the optics;
(v) excess radiative load in the dewar.
The noise terms add in quadrature, and the estimate for the 
relative contributions at the average sky emissivity for
our survey ($\left\langle\tau_{850}\right\rangle\,{\simeq}\,0.2$) 
are respectively 29, 25, 17, 13, and 16 per cent 
(W. Holland private communication).
Thus sky noise is the largest source, but does not dominate the others. 
Nevertheless, we expect the noise to integrate down with time.
This will not be true of one additional source of noise: `confusion noise'
arising from fluctuations due to undetected sources.  However, the level is
essentially negligible for these observations - Blain et al.~(1998) quote a
value of $\sigma_{\rm conf}\,{=}\,0.44\,$mJy, derived from their
source counts, while we find that any reasonable extrapolation of the
counts (together with the constraint of not over-producing the FIRB) leads
to a variance of no more than $0.5\,$mJy.  We can therefore safely ignore
confusion noise.  However, it is worth pointing out that lensing by the
clusters effectively allows us to probe below the confusion limit of say
$3\sigma_{\rm conf}$.

With transient noise and bad pixels from scan to scan, our
confidence in many of the detections is reinforced by dividing the
data into two parts and seeing evidence of the sources (at the ${>}\,2\sigma$
level) appear in both.
Multi-wavelength detections also substantially strengthen our 
confidence in a source. 
Although the 450\mum\ channel is typically not sensitive enough
to detect these objects, especially those lying at higher redshifts,
many more of them may be observable at 1350\mum\ with
deeper photometry than we have currently obtained.
Three sources have been subsequently verified through PHOTOMETRY followup
observations (MS\,0451-A, Abell\,2261-A, and MS\,1455-A). These are discussed
in the individual source section (4.2).

Empirically, the localized noise levels are seen to integrate down in our maps 
roughly as $t^{-1/2}$ (see also Ivison et al.~1998a). 
We note that in a typical 850\mum\ SCUBA field of 5.2 square arcmin with
${\sim}\,100$ independent beams we expect only ${\sim}\,0.1$ spurious 3$\sigma$
peaks from Gaussian noise, and and $\ll$\,1 spurious 4$\sigma$ peak from our
whole survey.
 
The completeness of a set of detections has typically been established using
synthetic 
source recovery techniques in Monte Carlo simulations (Smail \etal 1997, 
Hughes \etal 1998).
Our 3$\sigma$ and
source catalogues corresponds approximately to the 80 per cent
completeness level found in Smail et al.~(1997).  We estimate that our
4$\sigma$ catalogue is more than 90 per cent complete.

Our sub-mm observations of the clusters are presented in Tables~1 and 2. 
Table~1 lists the clusters, coordinates, and redshifts, along
with total exposure time (based on the number of integrations).
The last four columns present the average 850\mum\ and 450\mum\ sensitivities 
achieved after subtracting sources ${>}\,3\sigma$ from the field,  
and the number of
sources detected at greater than the 3$\sigma$ and 4$\sigma$ levels.
Table~2 lists the properties of the detected sources, 
with coordinates and observed flux densities in each of the three sub-mm
wavelength bands.  These sub-mm detections and upper limits 
(upper limits are given as 95 per cent Bayesian confidence regions, assuming
Gaussian errors and neglecting the negative flux region)
at 1350, 850, and 450\mum\, 
provide rough photometric estimates of the source redshifts (see e.g.~Hughes 
et al.~1998). However, the addition of radio data significantly increases  
confidence in these photometric redshifts.
The final two columns list the VLA radio flux limits, 
and estimates of the redshift from the sub-mm and
radio data, as described below.

%
%
\begin{table*}
{\scriptsize
\begin{center}
\centerline{Table~1}
\vspace{0.1cm}
\centerline{Sub-mm observations for the clusters in our sample}
\vspace{0.1cm}
\begin{tabular}{lcccccccc}
\noalign{\smallskip}
{Cluster} & {R.A.$^{\rm a}$} & {Dec.$^{\rm a}$} & {$z$} & {Exposure time} &
{$S^{\rm b}_{850}(1\sigma)$} & {$S^{\rm b}_{450}(1\sigma)$} & 
\multispan2{\hfil $N(S_{850}\,{>}\,S_{\rm lim})$ \hfil } \cr
{} & (J2000) & (J2000) & {} & {(ks)} & {(mJy)} & {(mJy)} & {3$\sigma$}
 & {4$\sigma$}  \\
\hline
\noalign{\smallskip}
Cl\,0016+16 & 00~18~33.2 & $+$16~26~18 & 0.541 & 20.5 & 3.1 & 36  & 2 & 0  \cr
MS\,0451-03 & 04~54~13.4 & $-$03~01~47 & 0.550 & 14.1  & 4.2 & 181 & 2 & 2 \\
Abell\,520 & 04~54~19.0 & $+$02~56~49 & 0.203 & 7.7      & 5.0 & 66  & 2 & 1 \cr
Zwicky\,3146 & 10~23~39.6 & $+$04~11~10 & 0.291 & 12.8 & 2.6 & 25  & 4 & 1 \cr
MS\,1054-03 & 10~57~00.2 & $-$03~37~27 & 0.833 & 7.7 & 4.3 & 103 & 1 & 0 \cr
MS\,1455+22 & 14~57~15.1 & $+$22~20~35 & 0.259& 15.4& 2.1 & 22  & 1 & 1\cr
Abell\,2163 & 16~15~34.1 & $-$06~07~26 & 0.201 & 7.7   & 5.8 & 44  & 0 & 0 \cr
Abell\,2219 & 16~40~20.5 & $+$46~42~59 & 0.228 & 15.4  & 1.7 & 8   & 4 & 1  \cr
Abell\,2261 & 17~22~24.1 & $+$32~07~45 & 0.224 & 7.7   & 3.8 & 105 & 1 & 1 \cr
\noalign{\smallskip}
{\bf Total} & & & & & & & {\bf 17} & {\bf 7}\\
\noalign{\smallskip}
\noalign{\hrule}
\end{tabular}
\end{center}
\begin{flushleft}
{$^{\rm a}$ x-ray centre -- {\sl ROSAT} archive.}\\
{$^{\rm b}$ Measured RMS over the central 60 arcsec of the field after
subtracting a scaled SCUBA beam profile from any identified sources in
the region.}
\end{flushleft}
}
\end{table*}

%
%
\begin{table*}
{\scriptsize
\begin{center}
\centerline{Table~2}
\vspace{0.1cm}
\centerline{Details of individual sources}
\vspace{0.1cm}
\begin{tabular}{lcccccc}
\noalign{\smallskip}
Source & 
$S^{\rm b}_{1350\mu m}$
 & $S^{\rm b}_{850\mu{\rm m}}$ & $S^{\rm b}_{450\mu m}$ &
$S^{\rm c}_{\rm 1.4\,GHz}$ & $z^{\rm d}_{\rm est}$ \cr
{} & 
	& (mJy) & (mJy) & (mJy) & ($\mu$Jy) & {} \cr
\hline
\noalign{\smallskip}
Cl\,0016 & & & &  \cr
$\>\>$A\ SMMJ$\,00186{+}1626$
	& & 9.9$\pm$3.3 & $\phm\pho7\pm34\pho$ & 
	$<$200 & $>$1.8  \cr
$\>\>$B\ SMMJ$\,00186{+}1625$
	& & \pho9.8$\pm$3.2 & $-\pho4\pm36\pho$ &  
	$<$200 & $>$1.8  \cr
MS\,0451 & & & &  \cr
	$\>\>$A\ SMMJ$\,04541{-}0302$
	& & 16.8$\pm$4.2 	
	& $\phm24\pm181$ & $<$200 & $>$2.2 \cr
	$\>\>$B\ SMMJ$\,04542{-}0301^{\rm g}$
	& & 19.1$\pm$4.2 & $-61\pm177$ 
	& $<$200 & $>$2.3 \cr
Abell\,520: & & & &  \cr
	$\>\>$A\ SMMJ$\,04543{+}0257$
	& & 33.0$\pm$5.0 & $-39\pm67\pho$ &  $<$420
	& $>$2.2 \cr
	$\>\>$B\ SMMJ$\,04543{+}0256$
	& & 16.2$\pm$5.1 & $\phm13\pm70\pho$ &  $<$410
	& $>$1.6 \cr
Zwicky\,3146 & & & &  \cr
	$\>\>$A\ SMMJ$\,10237{+}0411$ 
	& $1.7\pm3.9$ 
	& \pho9.3$\pm$2.5 & $\phm15\pm23\pho$& 
	482/2722$^{\rm e}$ & 1.1(0.3)$^{\rm e}$  \cr
	$\>\>$B\ SMMJ$\,10237{+}0410$
	& & \pho9.2$\pm$2.5 &$-\pho2\pm25\pho$ &  543& 1.1  \cr
	$\>\>$C\ SMMJ$\,10237{+}0412$
	& & \pho9.4$\pm$2.6 &$\phm\pho6\pm27\pho$ 
	& $<$450 & $>$1.2 \cr
	$\>\>$D\ SMMJ$\,10236{+}0412$
	& & 11.2$\pm$2.6 
	&$\phm23\pm27\pho$ &  467& 1.3  \cr
MS\,1054 & & & &  \cr
	$\>\>$A\ SMMJ$\,10571{-}0337$
	& 15.2$\pm$5.4 & 14.8$\pm$4.3 
	&$\phm12\pm103$ & $<$200/1600 & 2.1(0.8) \cr
MS\,1455: & & & &\cr
	$\>\>$A\ SMMJ$\,14573{+}2220^{\rm g}$
	& & 10.1$\pm$2.2  & $\phm\pho5\pm22\pho$ &  $<$294 & $>$1.5\cr
Abell\,2219 & & & &  \cr
	$\>\>$A\ SMMJ$\,16403{+}46440$
	& 4.3$\pm$2.0 & 10.6$\pm$1.7 & $\phm34\pm11\pho$ &  $<$220 &  $>$1.8 \cr
	$\>\>$B\ SMMJ$\,16403{+}46437$
	& & \pho5.8$\pm$1.7 & $-\pho2\pm8\pho\pho$ &  593 & 0.8  \cr
	$\>\>$C\ SMMJ$\,16404{+}4643$
	& & $<$5.7  & $\phm26\pm8\pho\pho$ & 280/9100 & 0.9(0.2)  \cr
	$\>\>$D\ SMMJ$\,16404{+}4644$
	& & \pho6.3$\pm$2.0  & n/a$^{\rm f}$ & 1270 & 0.4 \cr
Abell\,2261 & & & &  \cr
	$\>\>$A\ SMMJ$\,17223{+}3207^{\rm g}$
	& & 17.6$\pm$3.9 & $-56\pm105$ 
	& $<$452 & $>$1.6  \cr
\noalign{\smallskip}
\noalign{\hrule}
\noalign{\smallskip}
\end{tabular}
\end{center}
\begin{flushleft}
{$^{\rm a}$ Positional errors for SCUBA sources with identifications
at radio and optical wavelengths (e.g.~Ivison et al.~1998b,
Frayer et al.~1999) are ${\sim}\,6$ arcsec for 4$\sigma$ detections and
${\sim}\,8$ arcsec for 3$\sigma$ detections.}\\
{$^{\rm b}$ Estimated flux densities with $1\sigma$ errors.
All non-detection upper limits are given as 
95 per cent Bayesian confidence regions, assuming
Gaussian errors and neglecting the negative flux region}\\
{$^{\rm c}$ Bayesian 95 per cent limits at 1.4\,GHz for sources
in A\,520, Zw\,3146, MS\,1455, A\,2219, and A\,2261, have been extrapolated
from the value at 4.9\,GHz, using a spectral index $\alpha\,{=}\,0.8$
appropriate for star formation driven synchrotron emission.
For the data from Stocke et al.~(1999), only
3$\sigma$ upper limits are given.}\\
{$^{\rm d}$ Redshift estimate using Carilli \& Yun (1999) radio/far-IR 
correlation.}\\
{$^{\rm e}$ In cases where there is ambiguity between the central cluster
galaxy and a background galaxy, the radio flux of the cluster galaxy is also
listed, and the cluster redshift is given in brackets.}\\
{$^{\rm f}$ This source falls at the very edge of the short wavelength
array.}\\
{$^{\rm g}$ These sources were verified with 850/450\mum\ photometry. See
text for details.}
\end{flushleft}
}
\end{table*}

\section{Properties of individual sources}

\subsection{VLA radio counterparts}

Radio counterparts to sub-mm sources are expected due to the
synchrotron emission resulting from either an 
AGN or shocked gas in supernovae remnants from a starburst (e.g. Richards et
al.~1999).
Given the small number density of radio emitting objects, the chances of
finding a 1.4\,GHz VLA object of ${>}\,200\,\mu$Jy in our error circle by
chance is much less than 1 per cent (Richards et al.~1998).
By contrast, identifying optical counterparts to the SCUBA-selected galaxies 
is a non-trivial matter for two main reasons:
1) the beamsize at 850\mum\ (the optimal wavelength for these studies) is
${\sim}\,15$ arcsec, with pointing errors for the telescope of order
2$\,$arcsec; 2) the large, negative K-corrections (i.e.~increase in flux
density as the objects are
redshifted) of dusty star-forming galaxies at these wavelengths imply that
sub-mm observations can detect such objects at $z\,{>}\,1$ in an
almost distance-independent manner.
Thus several candidate optical galaxies are often present within the
uncertainty of the sub-mm detection, along with the possibility that the
actual counterpart is at much higher redshift and undetectable with
current optical imaging.
In this paper we restrict our attention to radio counterparts, and defer
detailed optical analysis of the sub-mm source positions to a separate
paper.

Carilli \& Yun (1999, 2000 -- CY) recently demonstrated that 
crude redshift estimates for distant dusty galaxies could be obtained in
the absence of an optical counterpart using
the spectral index between the sub-mm (850\,$\mu$m) and radio (1.4\,GHz)
wavebands ($\alpha^{850}_{1.4}$), with a mean galaxy model
parameterized as 
$z\,{=}\,0.050 - 0.308\alpha + 12.4\alpha^2 - 23.0\alpha^3 + 14.9\alpha^4$.
The model SEDs used by CY are based largely on lower redshift observations,
however they are able to successfully describe the few high-redshift galaxies 
which have multi-wavelength radio and sub-mm observations available.
Follow up studies by Blain (1999) have shown that
if lower dust temperatures are adopted for the
sub-mm population than are seen in the local sources used in CY's
models, then the allowed redshifts are slightly lower for a
given value of $\alpha_{1.4}^{850}$. 
The isolated cases which do not fall within the CY model predictions are
known to have strong radio-loud AGN components.
The important point is that the derived redshift provides a robust lower
limit, irrespective of the nature of the emission mechanism, AGN or
starburst, since an AGN contribution to the radio emission flattens the radio
spectrum. The modest scatter
between the models in CY suggests that the $\alpha^{850}_{1.4}$ 
technique currently provides the most 
useful limit on the redshifts of sub-mm galaxies, in the absence of an
optical counterpart.

Smail et al.~(2000) recently applied the CY analysis to deep radio
observations of the complete sample of SCUBA-selected galaxies from their
cluster lens survey (Smail et al.~1998, Ivison et al.~2000), 
revealing a higher redshift distribution
than previously expected from optical follow-up work.  
Using existing VLA data for our SCUBA cluster sample, we can thus
constrain the redshift
distribution of our population and set the stage for optical/near-IR 
follow-up identifications.

\begin{figure*}
\begin{minipage}{170mm}
\begin{center}
\epsfig{file=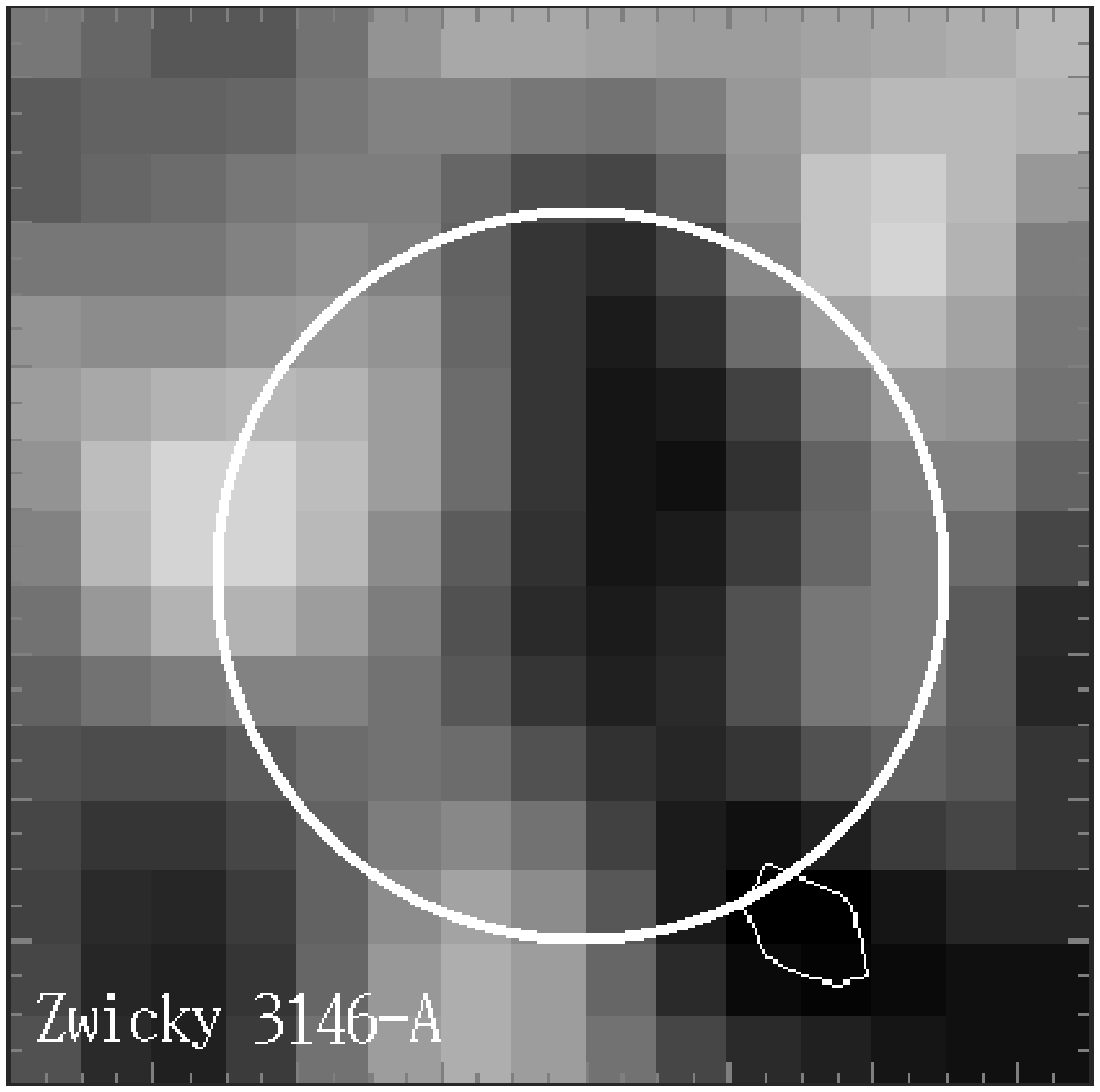, width=5.0cm, angle=0}
\epsfig{file=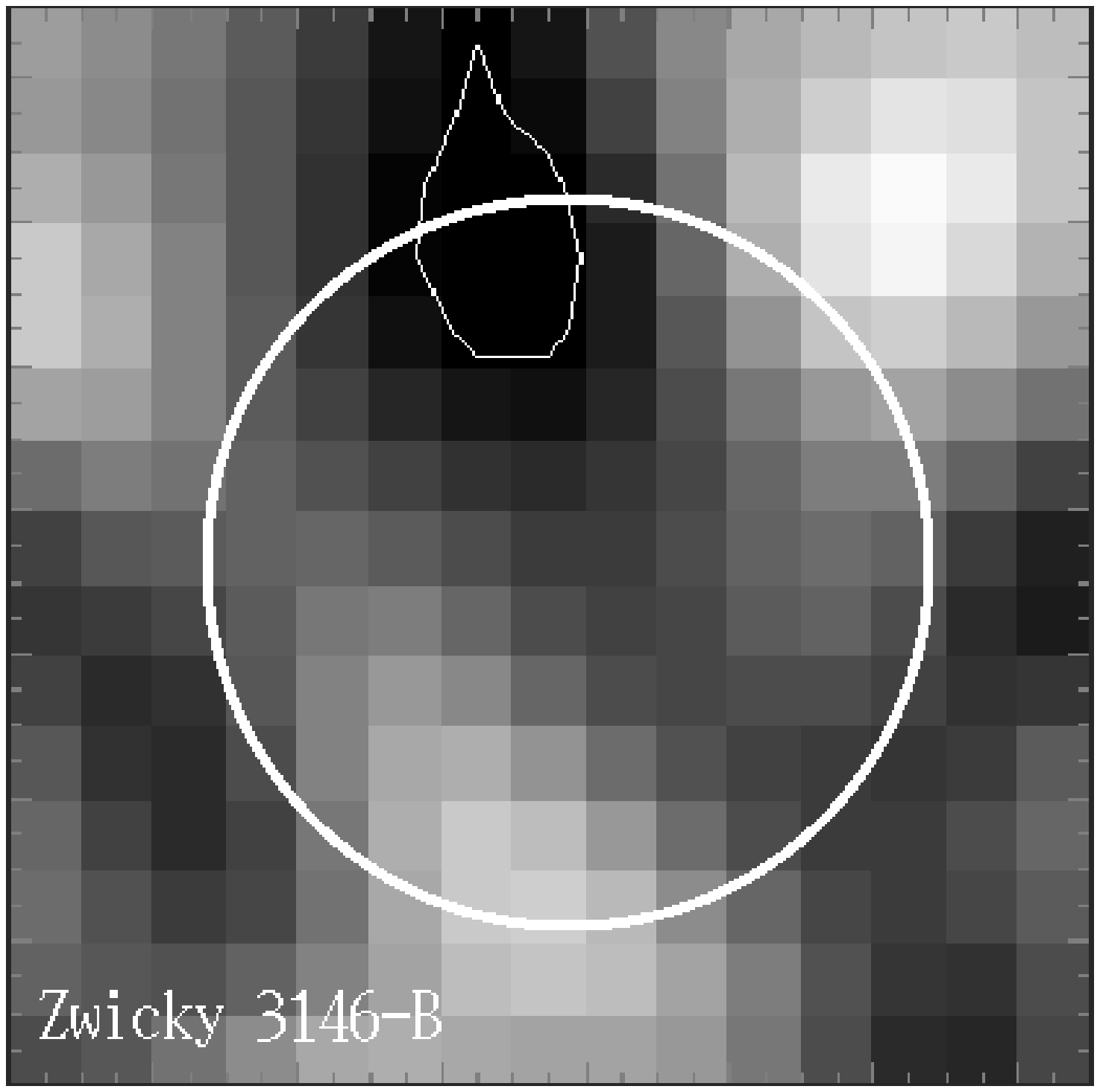, width=5.0cm, angle=0}
\epsfig{file=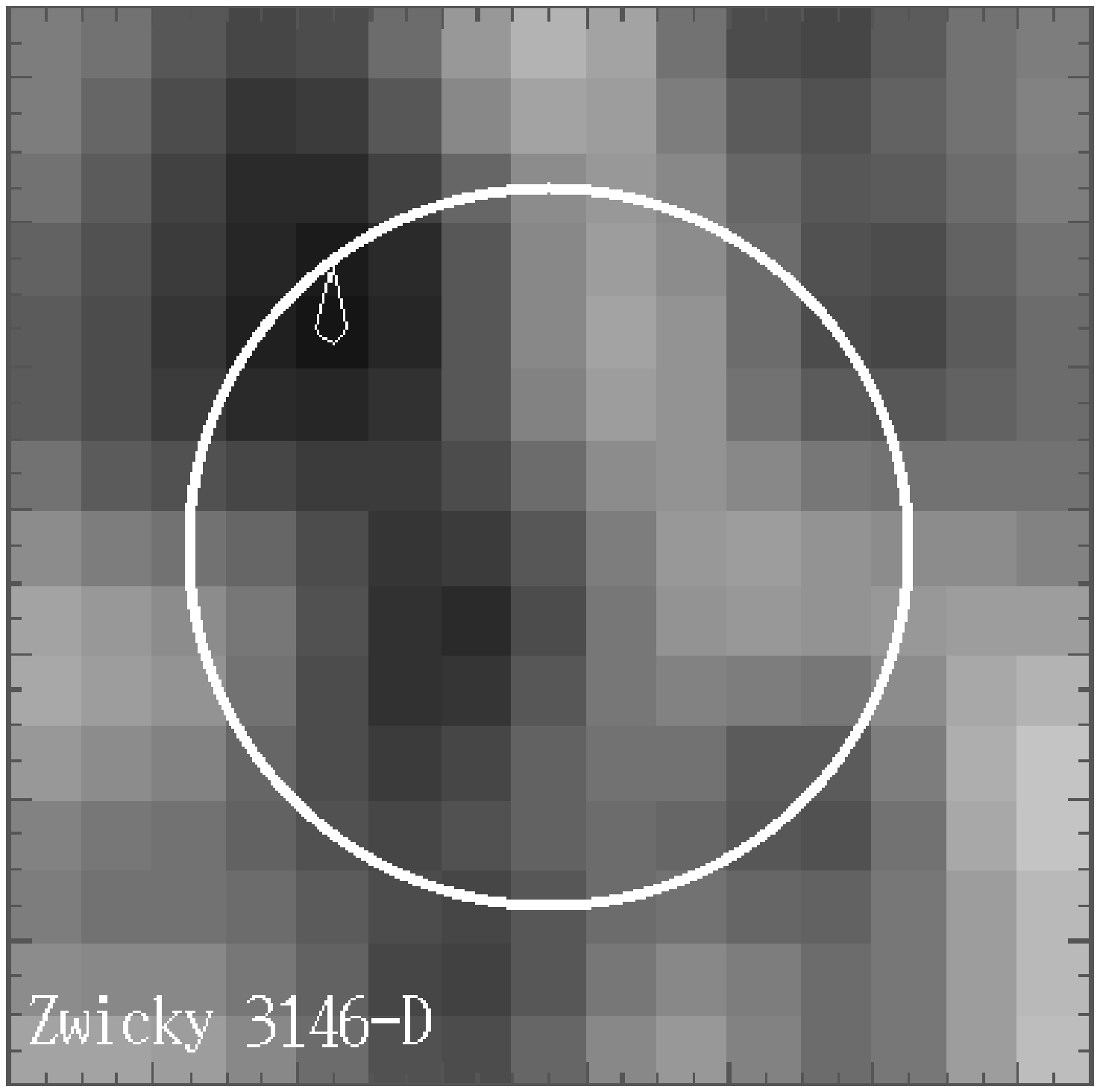, width=5.0cm, angle=0}
\epsfig{file=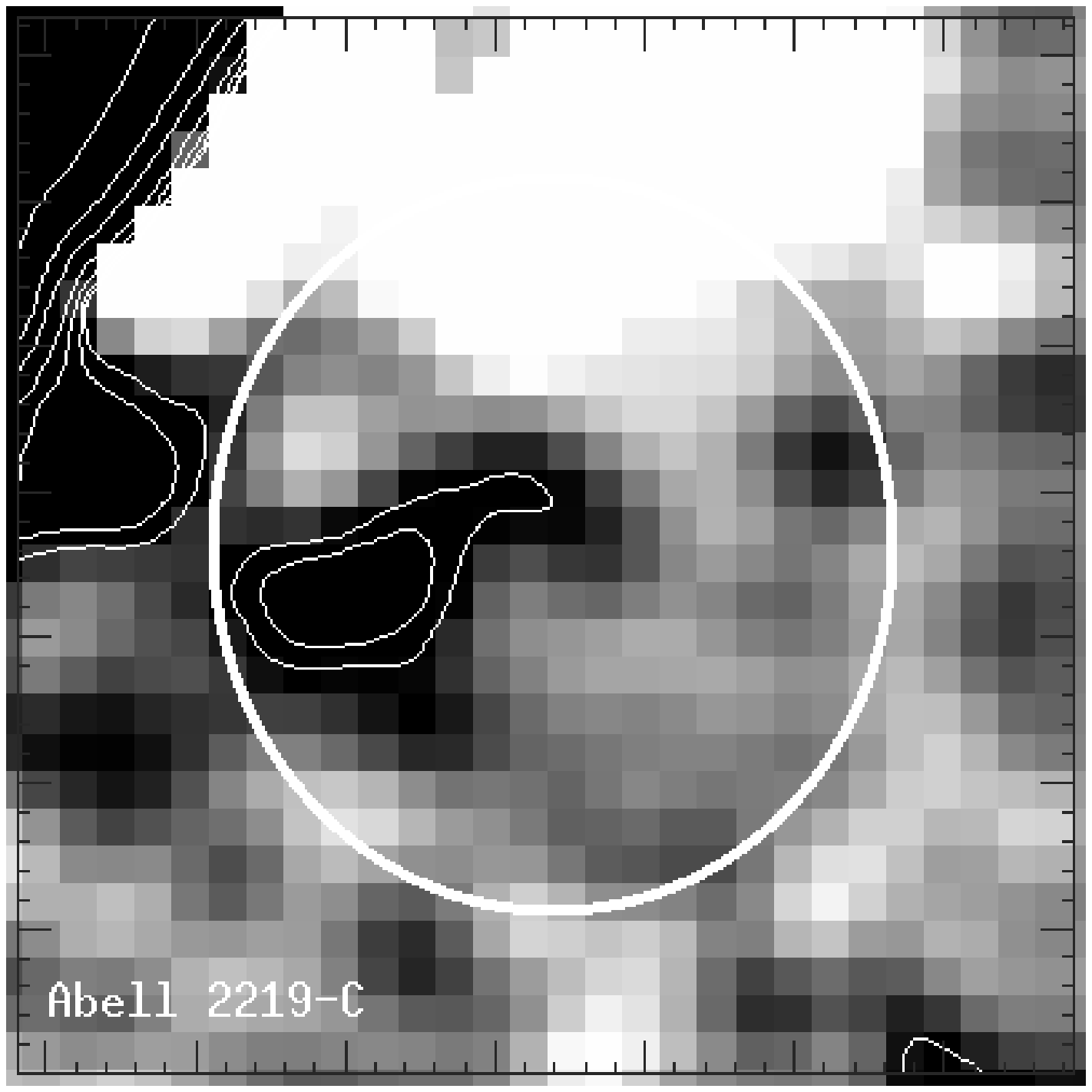, width=5.0cm, angle=0}
\epsfig{file=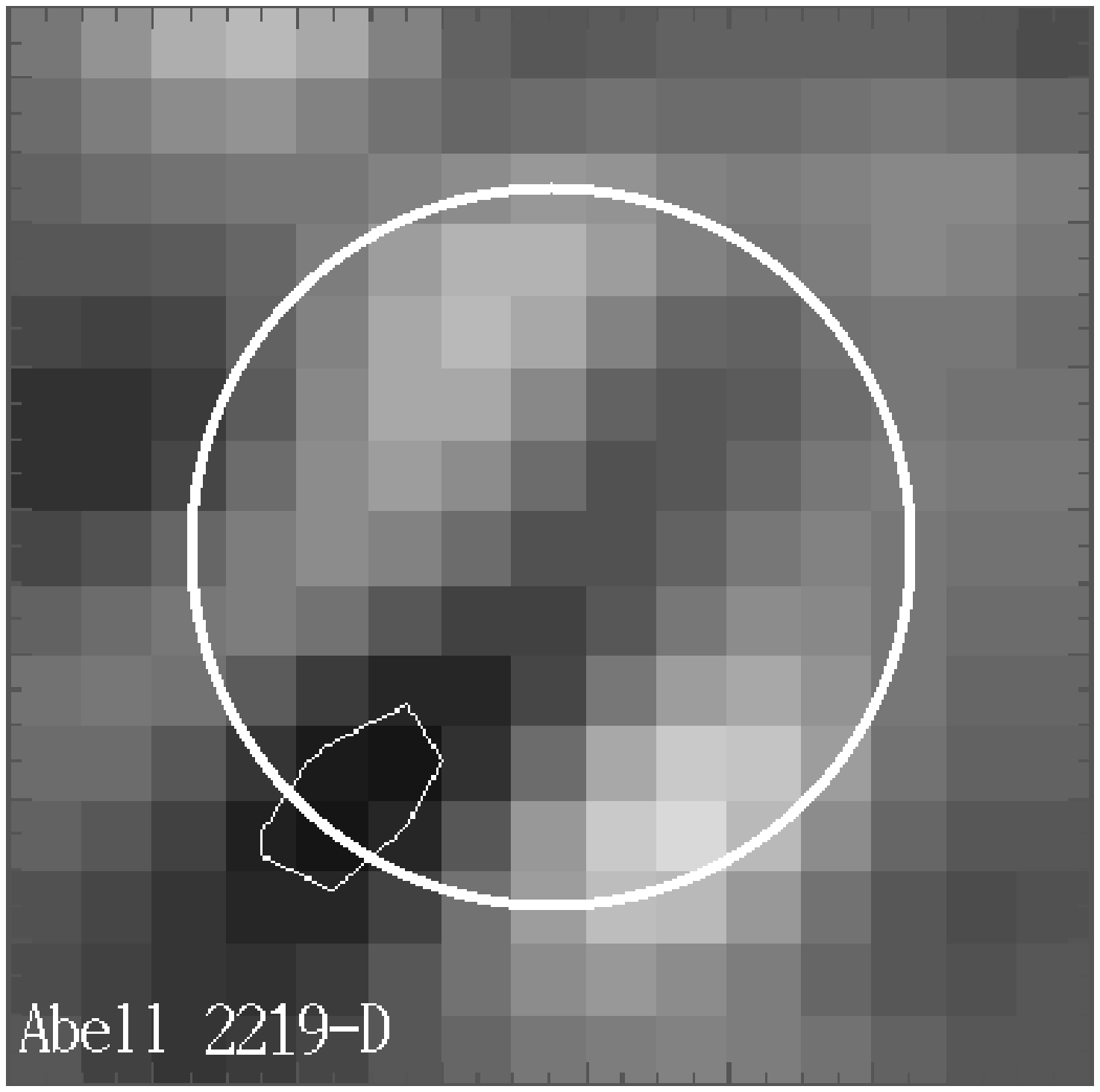, width=5.0cm, angle=0}
\end{center}
\caption{
VLA counterparts to our sub-mm detected galaxies.
We show VLA sources for cases where a ${>}\,3\sigma$ VLA
detection is seen relatively nearby the SCUBA position.
The radio contours are at 3, 5, 10, 20, 50, 100 and 200$\sigma$ in all cases. 
The circles are drawn at the positions of sources (see Table~2) from our
${>}\,3\sigma$ SCUBA source catalogue, with 10~arcsec diameter as an estimate
of the positional uncertainty.
The radio maps are at 4.9\,GHz except for Abell\,2219-C which is at 1.4\,GHz.
The other 12 SCUBA sources have only VLA upper limits, as listed in Table~2.
}
\label{vlacounterpart}
\end{minipage}
\end{figure*}

We searched for radio counterparts around the nominal positions of
the sub-mm sources based on the SCUBA astrometry.
Positional errors for SCUBA sources from the literature which were
clearly identified in radio and optical wavelengths (e.g. Ivison et al.~1998b,
Frayer et al.~1999) were ${\sim}\,6$ arcsec for 4$\sigma$ detections and
${\sim}\,8$ arcsec for 3$\sigma$ detections,
and we use similar error circles for our analysis.
Radio limits at 1.4\,GHz for sources in A\,520, Zw\,3146, MS\,1455, 
and A\,2261, have been extrapolated from the value at 4.9\,GHz, using a
spectral index $\alpha\,{=}\,0.8$ (i.e.~$S\,{\propto}\,\nu^{-0.8}$)
appropriate for star formation driven synchrotron emission. This
provides a reasonable estimate of the source redshift using the CY
relation (since a brighter radio counterpart leads to a lower redshift
estimate, while an AGN component would flatten the spectrum
giving rise to a higher redshift).
The 1.4\,GHz VLA data from Stocke et al.~(1999) are used to derive limits
for Cl\,0016, MS\,0451, and MS\,1054, and an archive 1.4\,GHz map of A\,2219 
(courtesy of F.~Owen) was used to measure fluxes directly. 
In 5 cases, there were radio sources ${>}\,3\sigma$ lying within the 
error circle of the sub-mm beam. One of these may be a central
cluster galaxy and is also discussed in the following section.
The 15 arcsec by 15 arcsec region surrounding the sub-mm
sources are  shown in Fig.~\ref{vlacounterpart}, with the
apparent radio flux densities or limits listed in Table~2.  

We use the CY relation (which is consistent with dust temperature 
$T_{\rm d}\,{=}\,50\,$K) to generate these lower limits to the source redshifts.
We note that the actual dust temperature in the sub-mm sources
may be lower, bringing down the
redshifts, or of course higher, thereby raising the inferred redshift.  In
general the constrained quantity is $T_{\rm d}/(1+z)$.

\subsection{Source characteristics and evidence for cluster members}
We now discuss each cluster field in detail:\\
{\bf Cl\,0016$\bf{+}$16}\\
The data for this cluster show a general positive excess at the
${\sim}\,2\sigma$
level, with 2 peaks barely reaching the $3\sigma$ level in the convolved map
(sources A and B).
There are no 1.4\,GHz VLA counterparts down to the 200\,$\mu$Jy level,
although the bright 2.7\,mJy radio souce from Stocke et al.~(1999) lies on
a clear sub-mm peak which reaches 2.6$\sigma$ significance
in the convolved map.
The generally positive excess could be caused by a Sunyaev-Zel'dovich 
increment (core radius ${\sim}\,2$ arcmin -- Carlstrom et al.~1999), 
which is too noisy in this map to detect with any
confidence.
However it may have some effect on the detectability of other sources in 
this field.\\ 
{\bf MS\,0451$\bf{-}$03}\\
An unresolved source (B: SMMJ$\,04542{-}0301$)
is found to the West in this cluster, and there is north-south 
extended positive
sub-mm emission over the x-ray centroid region, reaching 4$\sigma$ in the
convolved map (A: SMMJ$\,04541{-}0302$).  An additional
850\mum\ photometry observation centered on the elongated source
reveals a detection of 10.1$\pm$3.0\,mJy, which is consistent with the
peak flux densities found along the source as well as the positional accuracy
of the SCUBA map.  This central source is offset by 20~arcsecs to the east from 
the brightest cluster galaxy (BCG), and 20~arcsecs to the west from 
a 1.8\,mJy VLA source (Stocke et al.~1999). 
This detection could possibly represent cool dust associated with a cooling
flow or could be associated with the Sunyaev-Zel'dovich effect -- in either
case the emission would have to be very strong an isolated to a small region
of the cluster. For this particular cluster, the in-field chop reveals 
signs of the negative beams. However, another 
map of this cluster, observed with a large chop for
Sunyaev-Zel'dovich increment measurement (Halpern et al.~in preparation)
suggests that additional source structure in the central region would
significantly distort a clear offbeam signature.
The extended nature of this central source (also verified in our independent
map) is most likely to be a blend of two or more sources.  For the purposes
of Table~2 we treat this as a single source.  Positions and flux density
estimates for such blends should be treated with some caution until higher
resolution images are available.
\\
{\bf Abell\,520}\\
A large gradient appears to be present across the field, 
as evidenced by the
increased positive excursions in the lower halves of both the
450 and 850\mum\ images and negative excursions in the upper half. 
This could either be real or due to some residual atmospheric signal --
we could not find a convincing way to tell from our data, although we
suspect the latter explanation.
An apparent $4\sigma$ source (B: SMMJ$\,04543{+}0256$) lying to the
south of the primary source (A: SMMJ$\,04543{+}0257$),
is ${\sim}\,3\sigma$ above the noise even once a
linear gradient is fit to the plane and subtracted. The flux density
presented in Table~2 for this source reflects the subtraction of this
linear gradient component.
There is no detection of either the BCG or the
bright radio source lying to the east in the archive VLA map.\\
{\bf Zwicky\,3146}\\
The central source in Zw\,3146 (A: SMMJ$\,10237{+}0411$)
appears to be extended east-west, and may be another blend.
The VLA map reveals that the western extent
of the SCUBA source is associated with the BCG, with fairly bright radio
emission ($S_{\rm 1.4\,GHz}\,{=}\,0.92\,$mJy).
However, the eastern part of the SCUBA source shows no radio emission
(${<}\,210\,\mu$Jy), and the CY analysis
suggests a high redshift background source.
Sources B and C in this cluster field also have weak 4.9\,GHz detections
in the archive VLA map.\\
{\bf MS\,1054$\bf{-}$03}\\
The source in the MS\,1054 field (A: SMMJ$\,10571{-}0337$),
has a 1350/850\mum\ flux density ratio near unity. As thermal dust 
emission alone would produce a steep $\propto \nu^{3.5}$ spectrum,
this may indicate that the object  
has a flat-spectrum AGN radio contribution at 1350\mum,
and perhaps at 850\mum\ as well. 
A bright VLA counterpart, offset by 15 arcsec from the sub-mm peak, has been
identified by Condon et al.~(1998) as 
a cluster AGN radio source, and the CY relationship is consistent within
errors with the source lying in the cluster at $z\,{\sim}\,0.8$.
However, such a large radio/sub-mm offset is unlikely (see for example Richards
et al.~1999), and SMMJ$\,10571{-}0337$ could easily be a background galaxy
lensed by the cluster. Within errors, the 1350/850\mum\ ratio 
is not inconsistent with a higher redshift source. This source has been
verified in another SCUBA map of this cluster, although at a slightly lower
flux level (P.~van der Werf, private communication).\\
{\bf MS\,1455$\bf{+}$22}\\
There is some evidence for 850\mum\ SCUBA flux associated with the
BCG in the raw MS\,1455 maps. 
However, a somewhat noisy central pixel during the observations
of this cluster skews the peak profiles sufficiently that our source
finding routine does not identify this as a detection (the source
appears at only 2.6$\sigma$ in the final convolved map). We do not include
this as a source in our catalogue until it can be verified.
However, the proximity of the sub-mm peak to 2 bright radio sources 
(Fig.~\ref{coolingflow}), and a possible CO(1-0) detection for the BCG
(Edge et al.~in preparation) lends credence to this putative source.
The bright sub-mm source to the south (A: SMMJ$\,14573{+}2220$)
has no VLA counterpart in the archive maps.
An additional 850\mum\ photometry observation centered on source-A
reveals a detection of 6.8$\pm$1.8\,mJy, which is consistent within the
relative errors between the mapping and photometry measurements, 
as well as the extended nature of this source and positional uncertainties.\\
{\bf Abell\,2163}\\
No sources were found in this cluster field down to a 3$\sigma$ sensitivity
of 17\,mJy (corresponding roughly to 6\,mJy after correction for lensing).
This cluster had the poorest sensitivity, and
we include it here for completeness only.\\
{\bf Abell\,2219}\\
This map has by far the lowest 450\mum\ rms, and as a result there are,
in addition to the 850\mum\ detections, two 450\mum\ detections at the
$3\sigma$ level.
The 1350/850\mum\ and 450/850\mum\ ratios and lack of a 4.9\,GHz
detection for source A (SMMJ$\,16403{+}46440$) are consistent
with the high redshift predicted from CY. 
This is the only source in A\,2219 that does not have at least positive
flux density in the VLA archive map.  The central source
(C: SMMJ$\,16404{+}4643$), is ${\simeq}\,2.2\sigma$ at 850\mum\, but has a 
fairly clear detection at 450\mum.  It does not quite align with the
central galaxy radio emission (see Fig.~\ref{coolingflow}), 
but within errors, it is possible that the source is a cluster member.
Indeed the large implied 450/850\mum\ flux density ratio is 
indicative of a rather low redshift source.  In the archive 1.4\,GHz map,
however, there is a $5\sigma$ source lying near the SCUBA centroid, offset
${\sim}\,7$ arcsec from the BCG.
Source D (SMMJ$\,16404{+}4644$) lies near the edge of the frame (and off the
450\mum\ frame entirely), and is only 3$\sigma$ above the
local noise level.  However the centroid is within
5 arcsec of an ISO 15\mum\  source (Barvainis, Antonucci \& Helou 1999),
with which the sub-mm source may be associated.\\
{\bf Abell\,2261}\\
Although the bright (${>}\,4\sigma$) source to the west 
(A: SMMJ$\,17223{+}3207$)
does not correspond with any radio
emission, the central 2.5$\sigma$ peak in the SCUBA map is identified
with a radio source coincident with the optical BCG (Fig.~\ref{coolingflow}).
An additional 850\mum\ photometry observation centered on source-A
reveals a detection of 13.6$\pm$2.8\,mJy, which is consistent within the
relative errors between the mapping and photometry measurements,
as well as the extended nature of this source and positional uncertainties.

\subsection{Cooling flows and centrally concentrated dust emission}

The high central densities of the x-ray gas in many luminous clusters
lead to predicted cooling times shorter than the Hubble time 
(e.g.~Fabian~1994).  This cooling gas, which
can not support the pressure exerted by the overlying hot gas,
is expected to flow into the cluster centre.  Supporting evidence for such
cooling flows is now available in the form of inverted x-ray temperature
profiles (e.g.~Buote 1999), and x-ray emission lines show
gas that has lost most of its thermal energy.
The fate of this cool gas remains controversial (e.g.~Braine \& Dupraz 1994)
as no sink has yet been clearly identified.

Observations of the central cluster galaxies in cooling-flow clusters
indicate that the cooling flows are not forming large numbers of
stars.  On the other hand, such stars could be hidden if the initial
mass function is biased towards low masses, making the
integrated stellar spectrum difficult to detect (e.g.~Fabian, Nulsen
\& Canizares 1982; Mathews \& Brighenti~1999).
However, the emission-line nebulosity around the central galaxies in
many cooling-flow clusters, seen on scales of up to
100\,kpc, have emission-line ratios suggestive of large amounts of dust
in the central regions -- this dust could
obscure any stars formed there (Hansen, J\"orgensen \&
N\"orgaard-Nielsen 1995; Allen et al.~1995).
{\sl HST\/} imaging of central cluster galaxies
(e.g.~for Abell\,1795 -- McNamara et al.~1996; Pinkney et al.~1996, for
Abell\,2597 -- Koekemoer et al.~1999)
has also directly detected dust lanes in some cases.
At present the total amount or temperature of this dust is
difficult to estimate from the scant observational evidence.
Whether the observed dust originate from the cooled
gas clouds themselves or from the on-going star formation remains unknown.

\begin{figure*}
\begin{minipage}{170mm}
\begin{center}
\epsfig{file=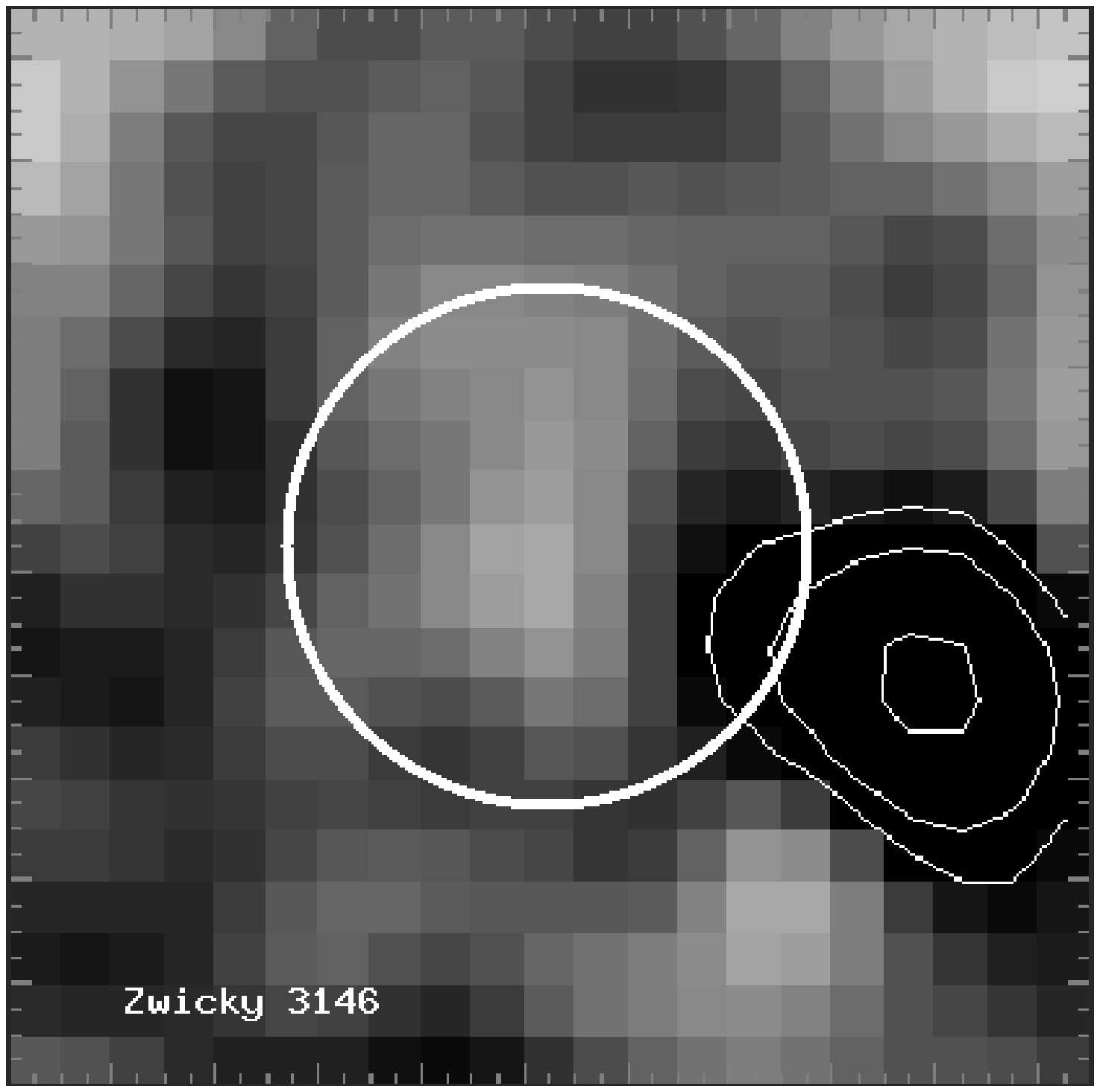, width=6.0cm, angle=0}
\epsfig{file=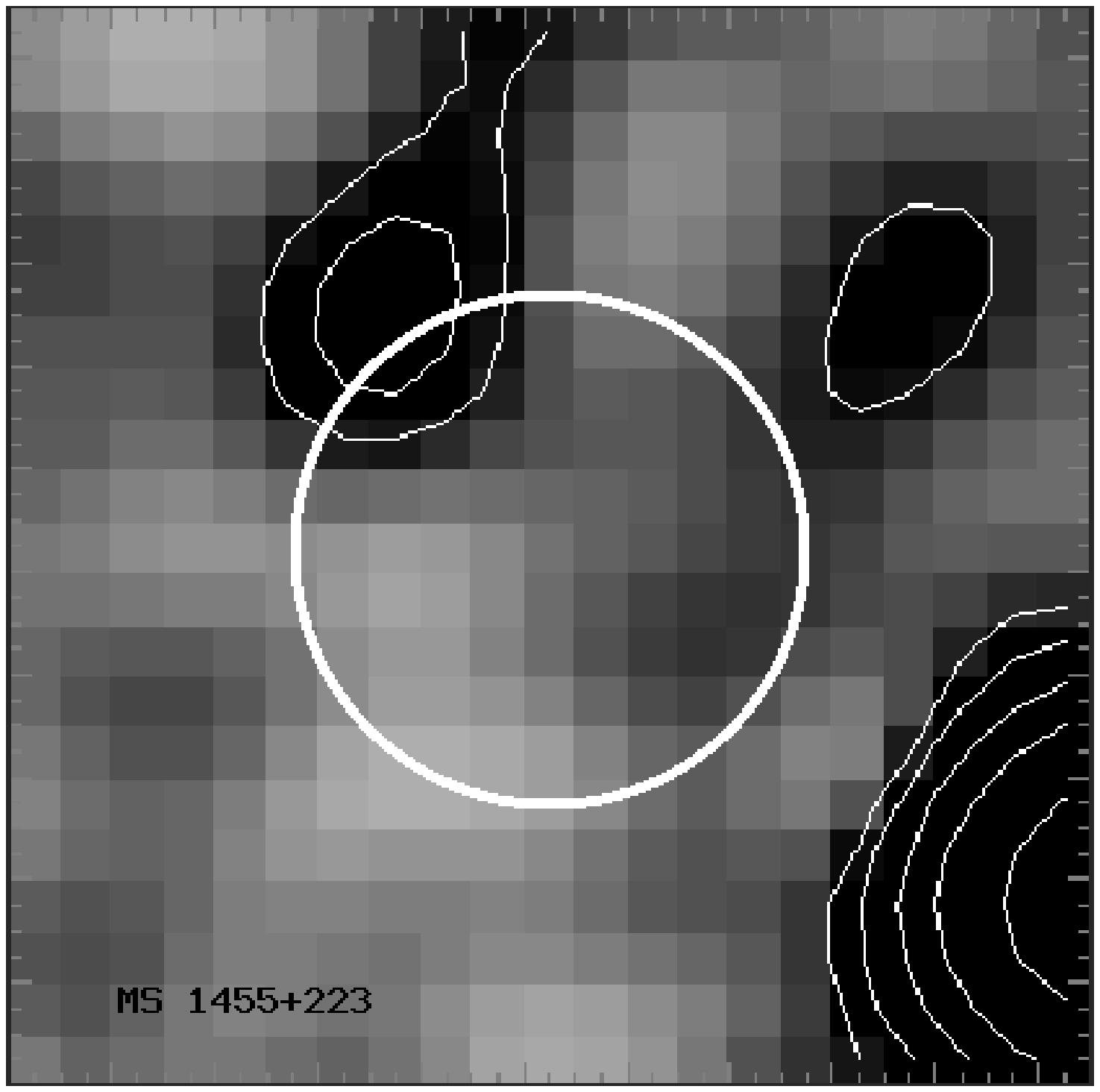, width=6.0cm, angle=0}
\epsfig{file=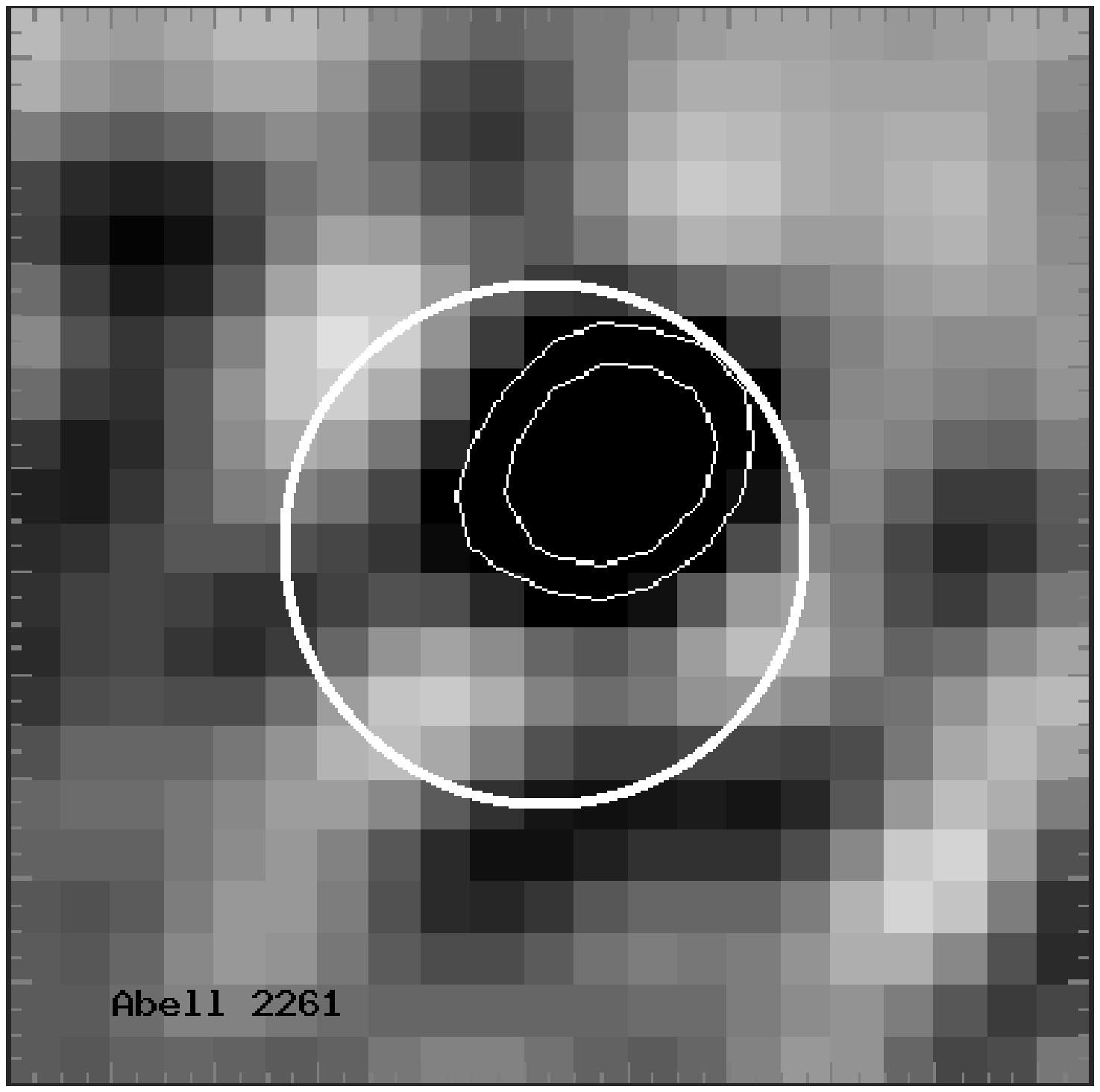, width=6.0cm, angle=0}
\epsfig{file=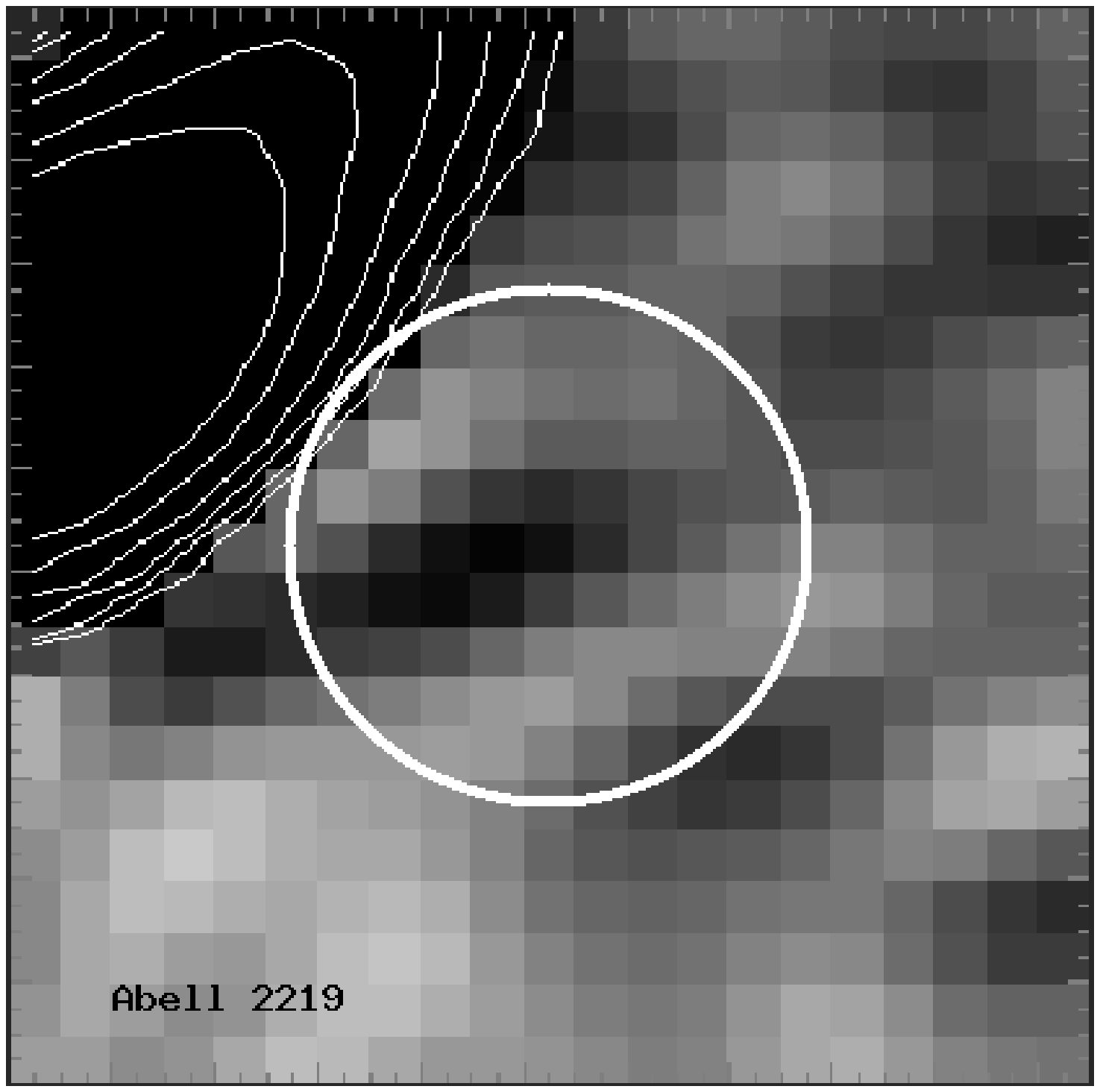, width=6.0cm, angle=0}
\end{center}
\caption{
VLA $4.9\,$GHz maps surrounding the brightest cluster galaxies
in Zw\,3146, MS\,1455, A\,2261, and A\,2219. The field size is
21 $\times$ 21~arcsecs in all cases. Radio contours are
3, 5, 10, 20, 50 and 100 times the RMS of the VLA map region.
A 10~arcsec diameter error circle is overlaid to indicate the SCUBA beam
position, for these marginal 2--3$\sigma$ 850\mum\ detections.}
\label{coolingflow}
\end{minipage}
\end{figure*}

SCUBA data may help here by probing emission from cool dust in or around
the central galaxies in clusters.
The implications of detected cool dust in the BCG of massive
clusters for the interpretation of star formation in
cooling-flow galaxies, has been discussed in Annis \& Jewitt (1993)
and Edge et al.~(1999)
Although there are no highly significant detections of the 
BCGs in our sample, 3 of the clusters show a marginal positive
flux density at 850\mum\ (${\sim}\,2.5\sigma$) near the BCG. In Abell\,2219,
a 3.5$\sigma$ detection at 450\mum\ (2.3$\sigma$ at 850\mum) lies
very near a radio luminous BCG (Fig.~\ref{coolingflow}). 
In addition, the cluster
MS\,0451 shows an extended 4$\sigma$ detection near the x-ray centroid
(Donahue~1996), although the brightest cluster galaxy is offset
from this position by 23 arcsec, and the only radio source in
the vicinity is offset to the other side by 20~arcsecs.
Radio maps of the 4 cluster centres with possible BCG detection,
along with 10 arcsec sub-mm error 
circles for reference, are shown in Fig.~\ref{coolingflow}.

%

Molecular gas, traced by CO emission, might also be present in strong
cooling flows (Braine et al.~1995).  For MS\,1455,
we also obtained an upper limit on the CO(4-3) emission from the BCG,
which suggests a gas-to-dust ratio of ${\la}\,150$ if the SCUBA peak were
actually a detection.  This would be somewhat unusual for known properties
of dusty BCGs (Edge et al.~1999), but a weak CO(1-0) detection for the
MS\,1455 BCG (A.~Edge, private communication)
lends support to a possible SCUBA detection.

If the SCUBA detection in MS\,0451 is associated with a cooling flow, then it
is fairly unusual, and may indicate a massive, but 
highly obscured optical counterpart
is actually being built by cooling flow fallout at the true centre of the
cluster potential.

%
%
\begin{table*}
{\scriptsize
\begin{center}
\centerline{Table~3}
\vspace{0.1cm}
\centerline{Limits on dust mass for central cooling flows.}

\vspace {0.1cm}
\begin{tabular}{lcccccc}
\noalign{\smallskip}
{Cluster} & {R.A.$^{\rm a}$} & {Dec.$^{\rm a}$} &  
{$S_{850}^{\rm b}$} & {$S_{450}^{\rm b}$} & {Dust mass$^{\rm c}$} & ${\rm M}_\odot$/yr$^{\rm d}$ \\
{} & {(J2000)} & {(J2000)}& {(mJy)} & {(mJy)} & {$h_{50}^{-2}{\rm M}_\odot$} \cr
\hline
\noalign{\smallskip}
CL\,0016+16 & 00~18~33.2 & $+$16~26~18  & $<$\pho6.1 & $<$\pho66 
& $<4.2\times10^8$ & n/a\cr
MS\,0451-03 & 04~54~13.4 & $-$03~01~47  &  19.1$\pm$4.2 & $<$354
 & $\phg1.3\times10^9$ & n/a \\
Abell\,520 & 04~54~19.0 & $+$02~56~49 &  $<$10.0 & $<$116
 & $<2.5\times10^8$ & $<$85 \cr
Zwicky\,3146 & 10~23~39.6 & $+$04~11~10  &  6.6$\pm$2.6 & $<$\pho48
 & $\phg2.4\times10^8$ & 1358\cr
MS\,1054-03 & 10~56~06.2 & $-$03~37~27  &  $<$\pho8.4 & $<$203
 & $<8.5\times10^8$ & n/a\cr
MS\,1455+22 & 14~57~15.1 & $+$22~20~35 &  5.3$\pm$2.1 & $<$\pho42
 & $\phg1.5\times10^8$ & 1227 \cr
Abell\,2163 & 16~15~34.1 & $-$06~07~26  &  $<$11.3 & $<$\pho86
 & $<3.1\times10^8$ & $<$90\cr
Abell\,2219 & 16~40~20.5 & $+$46~42~59  &  3.6$\pm$1.7 & $26\pm8$
 & $\phg9.0\times10^7$ & $<$485 \cr
Abell\,2261 & 17~22~24.1 & $+$32~07~45 & 9.5$\pm$3.8 & $<$205
 & $\phg1.8\times10^8$ & 680 \cr
\noalign{\smallskip}
\noalign{\hrule}
\end{tabular}
\par\noindent
\par\noindent
\end{center}
}
\begin{flushleft}
$^{\rm a}$ BCG position.\\
$^{\rm b}$ Where no positive value ${>}\,2.0\sigma$ is detected, we quote
Bayesian 95 per cent upper limits.\\
$^{\rm c}$ Assuming $T_{\rm d}\,{=}\,40\,$K and
$\beta\,{=}\,1.5$,
with $h_{50}=H_0/100\,{\rm km}\,{\rm s}^{-1}{\rm Mpc}^{-1}$.\\
$^{\rm d}$ Mass deposition values from Allen (2000). The three high-$z$
	clusters do not have the x-ray spatial resolution for 
	accurate deprojection (Donahue et al.~1999)\\
\end{flushleft}
\end{table*}
 
We can place limits on the dust mass for the BCG 
for all clusters in our sample, by considering the 850\mum\ flux density
limit at the redshift of the BCG (see Edge et al.~1999), and adopting model
dust parameters, for example
$T_{\rm d}\,{=}\,40\,$K, with an emissivity index of 1.5.
These limits along with the position of the BCG are presented in Table~3.
Corresponding CO(1-0) detections and limits, as well as broader implications
for cooling flows are presented in Edge et al.~(in preparation).
The limits are generally much less than the total mass deposition implied by
the cluster cooling flows.  But our observations are restricted to emission
around the BCG, and in any case the limits are model dependent.
Any conclusions we draw must necessarily not be very strong.
Our non-detections or relatively weak sub-mm emission from most BCGs in a
sample of nine clusters, imply that substantial effort will be required to
detect other central cluster galaxies.  And that if some of the mass is being
deposited as cool dust, this will be hard to detect
in typical central cluster galaxies with current instrumentation.

\section{Source Counts}
Source count models for rich cluster fields in the sub-mm have been studied
in detail by Blain (1997) and Blain et al.~(1999), with observational
constraints presented in Smail et al.~(1997) and Ivison et al.~(2000).  For
unmagnified blank fields, the   sub-mm galaxy counts can be derived by simply
dividing the number of detected sources by the surveyed area, although the
effects of clustering may have to be considered in small fields (Scott \&
White 1999).
The situation is more complicated in the case of lensing due to a massive
cluster, since the source plane is both distorted and magnified at a level
dictated by the source redshift and position.
Thus in the image plane, various regions are observed to different depths.
Although our clusters were not chosen to provide the largest
lensing amplification, the models of Blain (1997, figure~4) show that our
cluster redshifts are essentially optimal for maximum amplification,
with the exception of MS\,1054 which has slightly too high a redshift, but
still incurs a reasonably large lensing amplification. 

We model the amplification for each source and the effective depth of the
source plane area observed using an elliptical 
potential model of each cluster (Korman et al. 1994, Newbury 1997).
Parameters for the models are derived from both our own imaging data and 
from the best current information available in the literature. 
The lensing inversion is based on the prescription of Korman et al.~(1994)
and Newbury \& Fahlman (1999), allowing a
calculation of the lensing amplification and distortion as a function
of position in the image plane.
Table~4 presents the parameters for each cluster along with an estimate
for the lensing amplification at each source position.
Columns list the cluster name, redshift, cluster velocity dispersion,
angular distance and position angle from North
of each source relative to the x-ray centroid,
and resulting lens amplification. The final column lists the reference
for the cluster parameters.
All distances assume a flat $\Lambda$-dominated cosmology with
$\Omega_{\rm M}\,{=}\,0.3$.

The essential parameters of the model are the redshift of the cluster and
the 1-D (line-of-sight) velocity dispersion ($\sigma_V$), the latter being
essential for setting the mass scale.
These two quantities set the length scale in the image plane. 
Larger cluster redshifts reduce the effective magnification, while larger 
cluster velocity dispersions (cluster mass) increase the magnification.
Although more detailed mass models can in principle be constructed using the 
radial velocities of the individual cluster members (see Blain \etal 1999a),
the errors in the source counts are of order 30 per cent, and
our results would be very little affected by the use of such cluster
mass models.

There is still 
an ambiguity associated with the unknown background source redshift. However,
the models can be used to constrain the range of possible magnifications.
For sources at reasonably large redshift ($z\,{>}\,1$ say),
we find that the cumulative
source counts depend fairly weakly on the actual source redshift (less than
30 per cent scatter) in agreement with Blain \etal (1999a). If all sources are 
assumed to lie at $z\,{>}\,2$, the scatter reduces further. We assume
for simplicity that all sources lie at $z\,{=}\,2$.
Based on our CY radio analysis of the source redshifts, this is a 
reasonable assumption, although some sources may in fact lie at 
somewhat lower redshifts, especially if their dust temperatures are colder
than 50\,K.

%
\begin{figure}
\begin{center}
\psfig{file=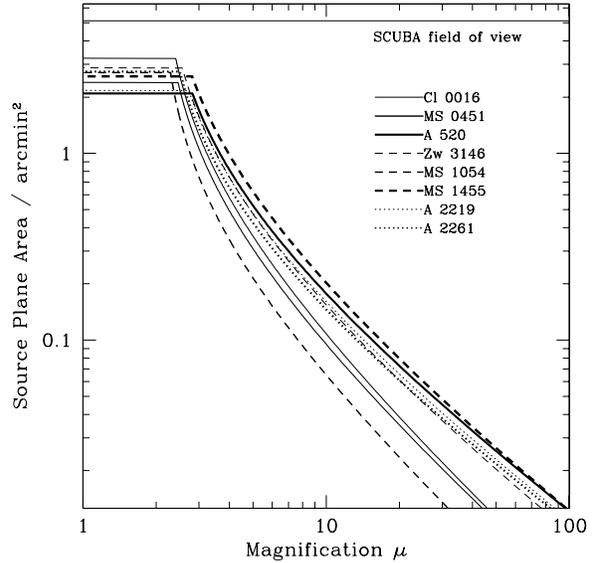,height=8truecm,angle=0}
\caption{Amplification vs area in the source plane for the 8 clusters in our 
sample with detected sources. 
}
\label{magarea}
\end{center}
\end{figure}

The cluster ellipticity ($E$) is chosen based on an outer contour for the
central galaxy, if one exists. In the absence of this BCG contour, we assume 
circular symmetry. The ellipticity is important only in the case of sources
near the cluster centre such as in Zw\,3146.
The core radius ($R_{\rm core}$) of the cluster is a model-dependent
parameter -- the core radius of the x-ray gas being temperature dependent. 
However for clusters with lensed arcs in their image, it is  
possible to estimate the core radius directly.
Again, the magnification is particularly sensitive to this
number only near the cluster centre, as it sets the location of the critical 
lines with very large magnification.
The ellipticity and core radius are quasi-independent parameters of the models.
For modest values of the ellipticity, this
parameter plays a very small role, especially in view of all the other
uncertainties involved.

Lensing increases the brightness of sources and also magnifies the area
-- so we reach to fainter effective flux limits, but over a smaller effective
solid angle.  Fig.~\ref{magarea} shows the resulting magnification vs
source plane area for each cluster. Our complete survey area in the source
plane is in fact only ${\sim}\,19$ arcmin$^2$, from a total surveyed 
SCUBA field area of ${\sim}\,47$ arcmin$^2$. We estimate that the
uncertainties in cluster parameters ($\sigma_V$, $R_{\rm core}$, $E$)
to the lensing model contribute ${\sim}\,25$ per cent error, a little larger
than the errors in calibrating SCUBA maps, but less than the Poisson error
in the source counts themselves.

%
%
\begin{table*}
{\scriptsize
\begin{center}
\centerline{Table~4}
\vspace{0.1cm}
\centerline{Lensing parameters for clusters}
\vspace{0.3cm}
\begin{tabular}{lccccl}
\noalign{\smallskip}
 {Cluster} & {$z$} & {$\sigma_{\rm V}({\rm km}\,{\rm s}^{-1}$} 
 & {$\Delta\theta_{\rm source}$ (PA)$^{\rm a}$} & Lens amplification
 & Cluster reference\cr 
\hline
\noalign{\smallskip}
Cl\,0016$+$16  & 0.541  & 1243 &   44(138), 62(171)  & 3.0, 2.6 &
Ellingson et al.~1998\cr
MS\,0451$-$03  & 0.550 & 1353  &   65(130), 25(46)  & 2.6, 4.5& 
Donahue~(1996)\cr
Abell\,520 & 0.203 & 927  &    61(104), 74(138)  & 3.4, 3.1&
Gioia \& Luppino~(1994)\cr
Zwicky\,3146 & 0.291 & 1310 &  31(117), 62(179), 69(-33), 78(-136)
	 & 6.8, 3.7, 3.5, 3.2 & Ebeling et al.~(1996)\cr
MS\,1054$-$03  & 0.833 & 1170 &  63(70) & 2.6 &
Donahue et al.~(1998)\cr
MS\,1455$+$22 & 0.259 & 1168 &  58(162.5)  & 2.8 & 
Stocke et al.~(1991)\cr
Abell\,2163 & 0.201 & 1231 & n/a  &n/a  & Gioia \& Luppino~(1994) \cr
Abell\,2219  & 0.228 & 1000
 &  66(17), 45(157), 15(178), 41(-93) & 3.6, 3.2, 7.5, 3.0 &
Smail et al.~(1996)\cr
Abell\,2261 & 0.224 & 1102 &  72(-124) & 3.3 & 
Ebeling et al.~(1996) \cr
\noalign{\smallskip}
\noalign{\hrule}
\noalign{\smallskip}
\end{tabular}
\end{center}
\begin{flushleft}
{{\bf$^{\rm a}$}
Distance and position angle of the sources (listed in Table~2)
from the x-ray centre.}
\end{flushleft}
}
\end{table*}

\section{Results}

%
%
\begin{figure}
\begin{center}
\psfig{file=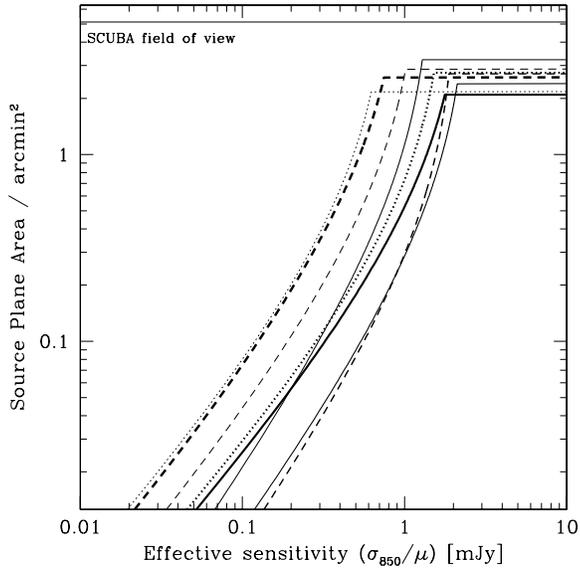,height=8truecm,angle=0}
\caption{Area vs sensitivity for the 8 clusters in our sample
with detected sources. Line types are the same as in Fig.~5.
}
\label{sensitivearea}
\end{center}
\end{figure}

\begin{figure}
\begin{center}
\psfig{file=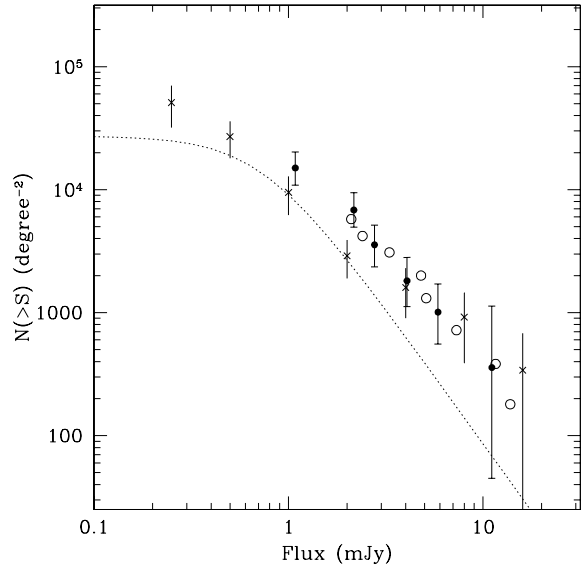,height=8truecm,angle=0}
\caption{The integral source counts 
(number of objects brighter than flux density $S$ as a
function of $S$.  We plot our total counts of ${>}\,3\sigma$ detections
(solid circles, binned), and our
sub-sample excluding possible cluster members and detections less than
4$\sigma$ significance (open circles, individual sources).
The corrected counts from Blain et al.~(1999c) are plotted with crosses.
The dotted line depicts a fit to the combined
blank field sub-mm counts (Hughes et al.~1998, Barger et al.~1999b,
Eales et al.~1999) constrained at faint fluxes
to not overproduce the far-IR background.
The error bars shown for our complete sample points are $\pm1\sigma$
(68 per cent Bayesian confidence region) errors based on Poisson statistics.}
\label{fig:counts}
\end{center}
\end{figure}

With the lensing factored in, our effective noise level in the central
regions of the cluster fields decreases sharply, as displayed in 
Fig.~\ref{sensitivearea}.
{}From these relations of sensitivity as a function of cluster area,
we can calculate the effective area surveyed to a given depth.
Fig.~\ref{fig:counts} shows the integral source counts from our survey 
including all sources detected at ${>}\,3\sigma$ (solid circles, binned).
We plot the counts from the Smail, Ivison, Blain (1997) lensing survey 
presented in Blain et al.~(1999c, crosses), as well as a fit to the combined 
blank field sub-mm counts (Hughes et al.~1998, Barger et al.~1999b, 
Eales et al.~1999) constrained at faint fluxes 
to not overproduce the far-IR background (dotted line).
When we restrict ourselves to the ${>}\,4\sigma$
sub-set which contains no obvious cluster members there is no dramatic
change in the counts (open circles in Fig.~\ref{fig:counts}).
We fit these $4\sigma$ counts with a simple power law 
$N({>}S)\,{=}\,N_0 (S/S_0)^{-\gamma}$, with $S_0=1\,$mJy,
and find $\gamma\,{\simeq}\,1.73$,
and $N_0\,{\simeq}\,11200$. These are close to the blank field values
of $\gamma\,{=}\,2.2$, and $N_0\,{=}\,13600$.
A direct comparison with other published sub-mm source counts thus 
reveals a good agreement between results, including gravitationally lensed and
blank field counts.  
SCUBA preferentially selects galaxies with $z\,{\ga}\,1$, due to
the spectral shape at these wavelengths. Therefore we expect few detections
in the clusters themselves, however, this may be the source of some of
the small discrepancy between the various surveys.

A positive magnification from the cluster lensing means that we are
seeing fainter galaxies, and hence an increase in the surface density of
galaxies.  However, this is offset by the loss in area 
due to field distortion. If the source count power-law index were too shallow,
this would result in a less efficient survey compared to 
a blank field survey with equivalent integration time.
However, our results show a net positive
bias, so that we are detecting more sources per unit time and
at unlensed flux densities below what would be possible in blank
field searches, in line with what was found earlier by
Smail et al.~(1997).

Using the amplification due to the foreground cluster, lensed surveys
also suffer less from confusion noise (e.g.~Blain et al.~1998).
The large beamsize of SCUBA implies a confusion limit at
${\sim}\,1.5\,$mJy, while we have detected 4 possible
sources fainter than 2\,mJy.

\section{Discussion}

Estimates of the redshift distribution for the SCUBA-selected sources
remain in dispute
(Hughes et al.~1998; Barger et al.~1999a; Lilly et al.~1999a).  Hughes et
al.~(1998) concluded that the bulk of the population is at
$z\,{\sim}\,2$--4, based on photometric redshift limits for the probable
counterparts of five sub-mm sources in the Hubble Deep Field
(c.f.~Richards 1999; Downes et al.~1999).  Barger et al.~(1999a)
undertook a spectroscopic survey of the Smail et al.~(1999) sub-mm sample 
and concluded that the median redshift was ${\sim}\,1.5$--2, with
the bulk of the population having $z\,{\sim}\,1$--3.  Lilly et
al.~(1999a) used archival spectroscopy and broad-band photometry of
sub-mm sources from the Eales et al.~(1999) survey to conclude that the
population spans $z\,{=}\,0.1$--3, with a third at $z\,{<}\,1$.  Given the
relatively small number statistics and the very real possibility of some
misidentifications, it is unclear whether there is any genuine
disagreement.

\begin{figure} 
\begin{center}
\psfig{file=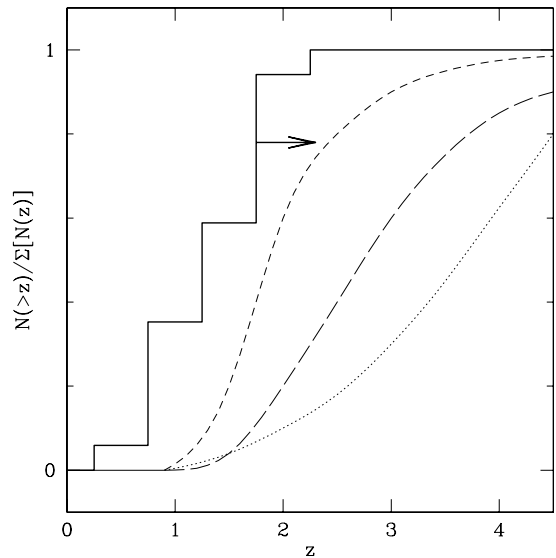,height=8truecm,angle=0}
\caption{Cumulative redshift distribution estimated
for our entire sample, derived from
the CY relation using VLA data (solid line). 
Also shown are the cumulative redshift distributions for  
models formulated to fit the far-IR background, and various infrared counts,
by Guiderdoni et al.~(1998) -- dotted line, Blain et al.~(1999b) Gaussian model
of luminosity evolution -- long-dashed line, and
Blain et al.~(1999b) modified Gaussian model fit to the Barger et al.~(1999a)
redshift distribution -- short-dashed line.
}
\label{redshifts}
\end{center}
\end{figure}

However, use of the Carilli \& Yun (CY -- 2000)
relation of radio/far-IR flux to predict the redshifts for
a large sample of SCUBA sources (Smail et al.~2000), has suggested that
the population may have a median value lying between $z\,{=}\,2.5$ and 3.
This is consistent with 
our CY analysis using VLA maps, which
result in a median value lying at $z\,{\ga}\,2.0$, although dependent on the
dust temperature and emissivity adopted.
Fig.~\ref{redshifts} shows an estimate of the cumulative redshift
distribution for our entire sample, derived from
the CY relation using radio detections and lower limits. 
Also depicted are the cumulative redshift distributions for
models formulated to fit the far-IR background, and various infrared counts,
by Guiderdoni et al.~(1998), and  Blain et al.~(1999b).

The archival radio maps for many of our cluster fields
are not deep enough to put strong constraints on the
redshift distribution of sub-mm sources, with few actual radio detections.
Given that the derived redshifts are generally lower limits, it is 
difficult to presently discriminate between 
predictions for the redshift distribution from various models  
(e.g. Blain et al.~(1999b) model Gaussian and modified Gaussian models
(long and short dashes) or the Guiderdoni et al.~(1998) model (dotted line)).
However, the radio/far-IR correlation indicates that the
bulk of the dust emission in the universe did not occur at redshifts
much below $z\,{=}\,2$.
The true redshift distribution remains to be determined.  Deep optical
and near-IR imaging should also uncover these objects' optical properties,
and resolve currently unanswered questions such as whether the galaxies would
be detectable through their Lyman break (Chapman et al.~2000), or whether
they have extremely red colours (Smail et al.~1999).

We can calculate lower limits to the far-IR background radiation
intensities from measured flux densities of resolved 4$\sigma$ sources
in our cluster fields. 
The contribution is the sum
$\Sigma(S/A)$, where $A$ is the area which has a sensitivity
necessary to see a
source of flux $S$, where lensing has been factored into the sum.
We find $9.1\,{\rm Jy}\,{\rm deg}^{-2}$ averaged over all 9 fields.
This corresponds to 20 per cent of the total
FIRB at 850\mum\ as measured by Fixsen et al.~(1998):
$44\,{\rm Jy}\,{\rm deg}^{-2}$.
The accuracy of these estimates however depends on
the lensing models and other details of the counts, together with
the remaining uncertainty in the FIRB measurement as these wavelengths.
If we take the simple power law model integrated up from
$1\,$mJy, then the background is $N_0\gamma/(\gamma-1)$.  Using our
slope from Fig.~\ref{fig:counts} accounts for more than half of the
background down to $1\,$mJy.

\section{Conclusions}

\begin{enumerate} 
\item We have detected 17 new sub-mm sources in the fields of 9 rich galaxy 
clusters. Comparison with VLA radio images suggests that most of these are 
being lensed by the cluster mass, rather than being in the clusters
themselves.
\item We see no strong evidence for cool dust emission around BCGs,
which might be related to cooling flows.
\item Our sub-mm source counts are in reasonable agreement with
existing sub-mm surveys, and extend the list of such sources available
for follow up at other wavelengths.
\item The redshift distribution has been loosely 
constrained using the radio-far-IR correlation,
and favours a relatively high redshift ($z\,{>}\,2$)
for the bulk of dust emission in the Universe.
\end{enumerate} 

\subsection*{ACKNOWLEDGEMENTS}

This work was supported by the Natural Sciences and Engineering
Research Council of Canada.
The James Clerk Maxwell Telescope is operated by
The Joint Astronomy Centre on behalf of the Particle Physics and
Astronomy Research Council of the United Kingdom, the Netherlands
Organisation for Scientific Research, and the National Research
Council of Canada.
We would like to acknowledge the staff at JCMT for facilitating
these observations. 
The VLA is run by NRAO and is operated by Associated
Universities Inc., under a cooperative agreement with the National
Science Foundation.

\end{document}